\documentclass[aps,prl,showpacs,twocolumn,amsmath,superscriptaddress]{revtex4-1}

\usepackage[svgnames]{xcolor}
\usepackage{color}

\usepackage{pifont}
\usepackage{epsfig,amssymb,amsmath,amsthm,amsfonts,amsbsy,mathrsfs,amscd}
\usepackage{graphicx}
\usepackage[normalem]{ulem}

\usepackage{subfigure}

\usepackage{empheq}
\usepackage{tikz}  

\usepackage{yfonts,dsfont}
\usepackage{physics}

\usepackage{floatrow} 

\newtheorem{theorem}{Theorem}

\usepackage{enumerate}

\usepackage[unicode=true,
 bookmarks=true,bookmarksnumbered=false,bookmarksopen=false,
 breaklinks=false,pdfborder={0 0 1},backref=false,colorlinks=true]{hyperref}
 
\hypersetup{
 linkcolor=magenta, urlcolor=blue, citecolor=blue, pdfstartview={FitH}, hyperfootnotes=false, unicode=true}

\usepackage{overpic} 

\usepackage{scalerel} 

\usepackage{longtable}
\usepackage{mathtools} 

\usepackage{siunitx}

\usepackage{wasysym}

\usepackage{soul} 

\input amssym.def

\usepackage{notes2bib} 

\begin{document}

\title{Anomalous Heat Transfer in Nonequilibrium Quantum Systems}

\smallskip
\author{Teng Ma}
\affiliation{BGI Research, Shenzhen 518083, China}
\affiliation{BGI Research, Wuhan 430074, China}
\affiliation{Beijing Academy of Quantum Information Sciences, Beijing 100193, China}

\author{Jing-Ning Zhang}
\affiliation{Beijing Academy of Quantum Information Sciences, Beijing 100193,  China}

\author{Yuan-Sheng Wang}
\affiliation{School of Systems Science, Beijing Normal University, Beijing 100875, China}

\author{Hong-Yi Xie}
\email{xiehy@baqis.ac.cn}
\email{hongyi.xie-1@ou.edu}
\affiliation{Beijing Academy of Quantum Information Sciences, Beijing 100193, China}
\affiliation{Department of Physics and Astronomy, Center for Quantum Research and Technology, University of Oklahoma, Norman, Oklahoma 73069, USA}

\author{Man-Hong Yung}
\email{yung@sustech.edu.cn}
\affiliation{Central Research Institute, 2012 Labs, Huawei Technologies, Shenzhen 518129, China}
\affiliation{Department of Physics, Southern University of Science and Technology, Shenzhen 518055, China}
\affiliation{Shenzhen Institute for Quantum Science and Engineering, Southern University of Science and Technology, Shenzhen 518055, China}

\begin{abstract}

Anomalous heat transfer (AHT), a process by which heat spontaneously flows from a cold system into a hot one, superficially contradicts the Clausius statement of the second law of thermodynamics. Here we provide a full classification of mechanisms of the AHT in nonequilibrium quantum systems from a quantum-information perspective. For initial states in local equilibrium, we find that the AHT can arise from three resources: initial correlation, intrasystem interaction, and intrasystem temperature inhomogeneity. In particular, for qubit systems, we prove that initial quantum coherence is necessary for AHT if the intersystem interactions are limited to the two-body type. We explicitly show the AHT dominated by each of the mechanisms in a three-qubit system. 
Our classification scheme may offer a guideline for developing high-efficiency quantum autonomous thermal machines. 

\end{abstract} 

\maketitle

\textbf{Introduction.}
The second law of thermodynamics, as a fundamental physical law of nature, establishes the direction of thermodynamic processes. 
Its Clausius statement claims that heat can only spontaneously flow from a hot object into a cold one~\cite{clausius1879mechanical}, and, entropy, a state function characterizing the randomness of the isolated joint system, must be nondecreasing in the process. 
This thermodynamic irreversibility has been popularly termed ``the arrow of time''~\cite{eddington1930nature}. 
However, in an absolutely isolated and finite-sized quantum system, the second law holds only under certain initial conditions, since the system evolution is unitary and reversible~\cite{lloyd1989PRA,vedral2016}. 

One commonly discussed restriction is that the hot and cold systems should be weakly correlated~\cite{PhysRevE.77.021110,jennings2010PRE,binder2018,bera2017NC,vedral2016}. This weak-correlation condition faithfully describes the usual circumstance between macroscopic objects in daily life~\cite{boltzmann2003relation,zeh2007}, as Boltzmann suggested for classical systems~\cite{boltzmann1970weitere,boltzmann1897hrn}.
However, correlation effects can be rather significant in generic quantum systems, and, in particular, there exist quantum correlations such as quantum entanglement~\cite{vedral1997PRL,horodecki2009RMP} and quantum discord \cite{henderson2001JPAMG,ollivier2001PRL,modi2012RMP} having no classical counterparts, which play an important role in energy flow \cite{sarovar2010NP,lloyd2017,ma2019PRA}. 
It has been proposed that when two quantum systems are strongly correlated, ``anomalous'' heat transfer (AHT), that is, heat spontaneously flows from the cold system into the hot, can occur, at the cost of consuming the intersystem correlation~\cite{binder2018,PhysRevE.77.021110,jennings2010PRE,jevtic2012PRL}. More recently, the correlation-induced AHT has been experimentally observed in nuclear spin systems~\cite{micadei2019NC} and further investigated in a wide range of systems from trapped ions~\cite{medinagonzalez2020PRA} to wormholes~\cite{xian2020PRR}. 
In addition, the AHT phenomenon has also been discussed in the context of symmetry-protected edge/surface states 
\cite{rivas2017SRa}, frustrated classical spin systems  
\cite{bauer2023}, and quenching dynamics of hot quantum gases 
\cite{gnezdilov2023PRA}.
 
In spite of enormous progress, a general and systematic framework for analyzing AHT in quantum systems is still lacking. 
In this Letter, we generically define the heat transfer problem, and develop a quantum-information classification of the AHT mechanisms based on the notion of local-equilibrium states~\cite{lebon2008,zwanzig2001nonequilibrium,intravaia2016PRL,levy2014EL}. 
We find that, in addition to the intersystem correlation, the nature of interactions, as well as the intrasystem correlation and temperature inhomogeneity, can give rise to the AHT. 
We further investigate qubit systems in which the forms of interactions are limited by the so-called heat transfer condition (HTC). 
In particular, we prove that classical correlation is not sufficient to induce AHT if the intersystem interactions are of two-body type. We calculate the heat transfer for a three-qubit model and show explicitly that each of the mechanisms can independently realize the AHT.

\begin{figure}
\centering
\subfigure{
\begin{minipage}[]{0.465\textwidth}
\begin{overpic}
[width=1\textwidth]{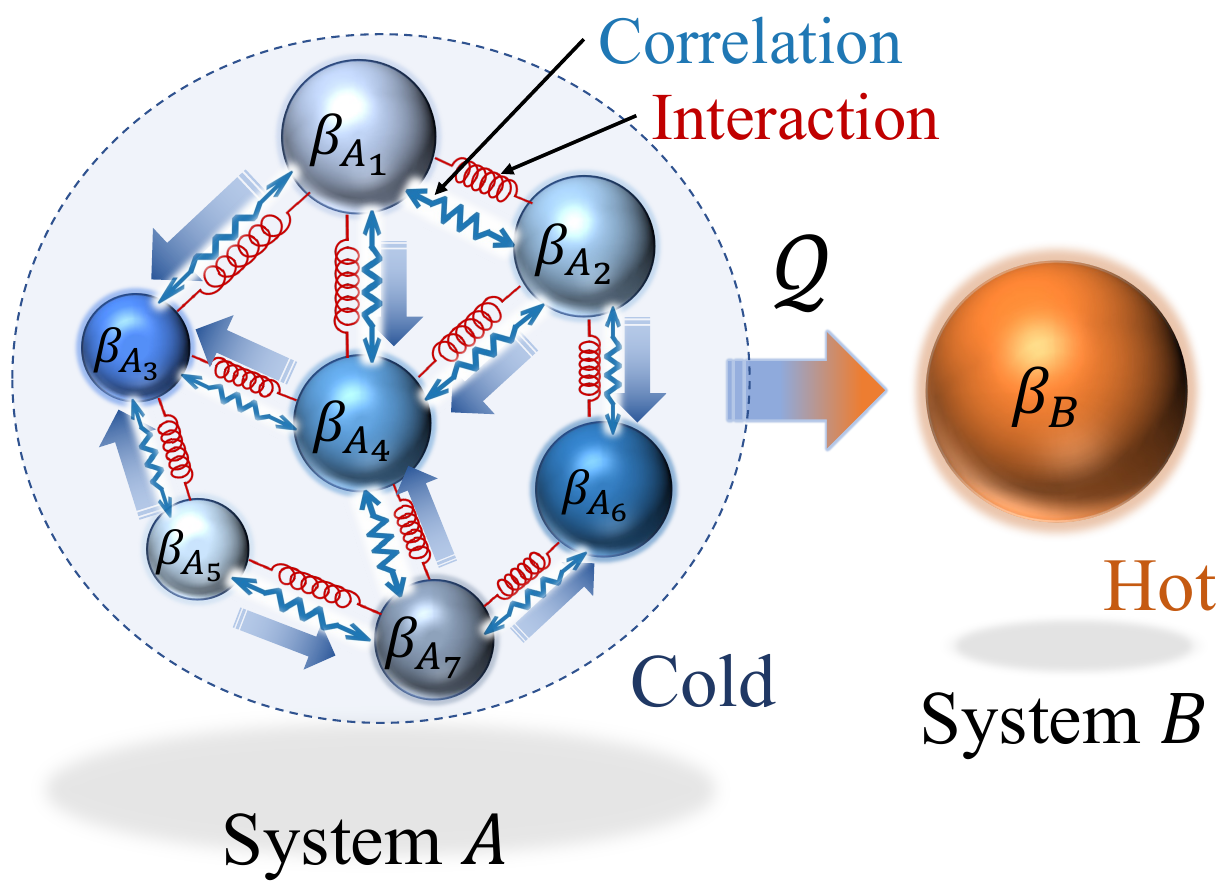}
\label{mulitibody_carton}
\put(0,63){\scriptsize{\textsf{(a)}}}
\end{overpic} 
\end{minipage}
}
\hspace{-0.05\textwidth}
\subfigure{
\begin{minipage}[]{0.5\textwidth}
\begin{overpic}
[height=0.8\textwidth]{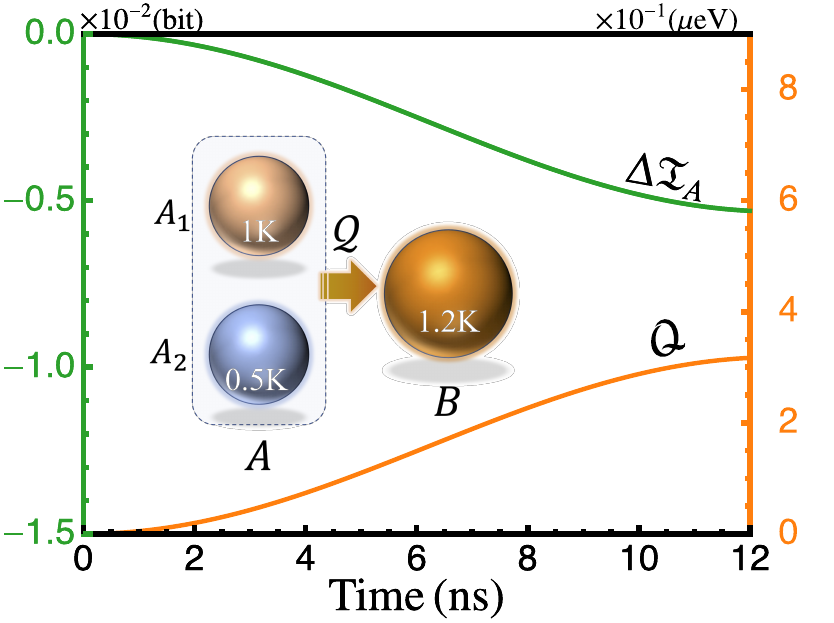}\label{ThreeQubits_InnerTemper}
\put(12,63){\scriptsize{\textsf{(b)}}}
\end{overpic} 
\end{minipage}
}

\vspace{-0.045\textwidth}
\hspace{-0.05\textwidth}
\subfigure{
\begin{minipage}[]{0.5\textwidth}
\begin{overpic}
[height=0.8\textwidth]{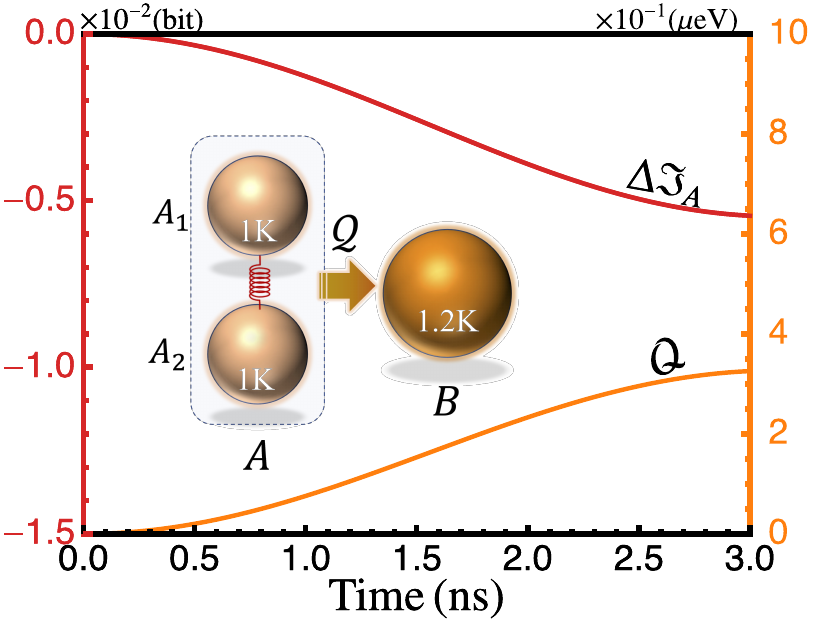} \label{ThreeQubits_InnerEnergy}
\put(12,63){\scriptsize{\textsf{(c)}}}
\end{overpic} 
\end{minipage}
}
\hspace{-0.05\textwidth}
\subfigure{
\begin{minipage}[]{0.5\textwidth}
\begin{overpic}[height=0.8\textwidth]{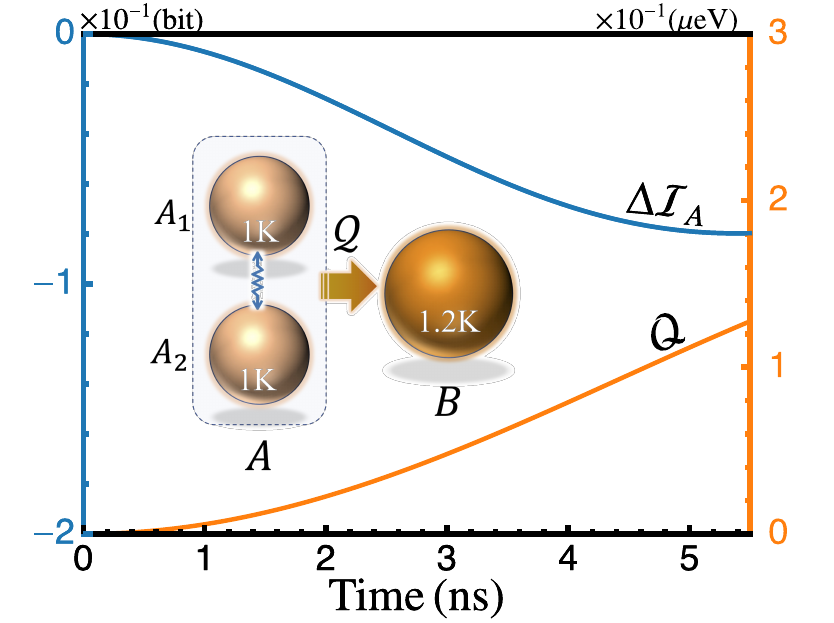}
\label{ThreeQubits_InnerCorre}
\put(8,63){ \scriptsize{ \textsf{(d)} } }
\end{overpic} 
\end{minipage}
}
\caption{AHT between two quantum systems that are initially in local equilibrium.
(a) Schematic diagram for the Hamiltonian in Eq.~\eqref{ham} and the initial LE state in Eq.~\eqref{les}.
The initial nonequilibration of the system $A$ is represented by temperature inhomogeneity ($\{\beta_{A_k}\}$), interactions (strings), and correlations (arrowed zigzags). We assume that $A$ and $B$ systems are the cold and the hot, respectively. 
(b)-(d) AHT in a three-qubit system. 
(b) Temperature-inhomogeneity-dominated AHT. We take the initial temperatures $T_{A_1}=\SI{1}{\K}$, $T_{A_2}=\SI{0.5}{\K}$, and $T_{B}=\SI{1.2}{\K}$, and the qubit frequencies $\omega_{A_2}=\omega_{B}= \omega_{A_1}/2=\omega_0$ with $\omega_0/2 \pi \equiv  \SI{4.0}{\GHz}$. 
(c) Interaction-dominated AHT. (d) Correlation-dominated AHT. In (c) and (d), we take $T_{A_1}=T_{A_2}=\SI{1}{\K}$, $T_{B}=\SI{1.2}{\K}$, and $\omega_{A_1}=\omega_{A_2}= \omega_{B}=\omega_0$.}
\label{AHT_examples}
\end{figure}


\textbf{Heat transfer Hamiltonian.} 
We consider heat transfer between two quantum systems $A$ and $B$, while the joint system is isolated. In general, the time-independent Hamiltonian of the joint system reads
\begin{equation} \label{ham} 
H=H_A + H_B+H_I,
\end{equation}
where $H_X$ with $X\in \{ A,B \}$ is the Hamiltonian of the system $X$ and $H_I$ describes the intersystem interactions. 
In Eq.~\eqref{ham}  we adopt the HTC, $[H_I, H_A + H_B]=0$, that is, the total energy of $A$ and $B$ is conserved, which ensures that the energy loss of $A$ equals to the energy gain  of $B$ during the evolution of the joint system 
\bibnote{If the HTC breaks, $H_I$ may induce energy variation in each of the systems, which has been discussed, for example, in Ref. \cite{gnezdilov2023PRA} (see also our discussions in Supplemental  Material).}. 
Therefore, as the joint system evolving from initial ($t=0$) state $\rho$ to finial state $\rho'$, one can unambiguously define the heat transfer $\mathcal{Q} \equiv \Delta \langle H_B \rangle=-\Delta\langle H_A \rangle$, where $ \Delta \langle O \rangle \equiv \tr [O (\rho' - \rho)]$ denotes the variation of the expectation value of an operator $O$. The two states $\rho$ and $\rho'$ are related by the evolution operation $\rho' = e^{-i H t} \rho e^{i H t}$, where we have taken $\hbar=1$. The HTC has been widely exploited in the study of thermal operations in the context of the resource theory~\cite{cwiklinski2015limitations,lostaglio2015NC,brandao2013resource,chitambar2019RMP}, and it tremendously constrains the parameters of the Hamiltonian. We note that, two joint systems, described by Hamiltonians $H_{1,2}$ and initial states $\rho_{1,2}$, exhibit identical heat transfer processes, if $U H_1 = H_2 U$ and $U\rho_1 = \rho_2 U$ with $U$ a time-independent unitary transformation.  

An instantaneous state $\varrho(t)$ satisfies the Liouville-von Neumann equation $d \varrho/dt = i [\varrho,H]$. In consequence, the $n_\mathrm{th}$ time derivative of the heat transfer can be written as 
\begin{equation} \label{n-d}
\frac{d^n \mathcal{Q}}{d t^n} = i^n \big\langle [H_I,H_B]_n \big\rangle_\varrho,
\end{equation} 
where the operator $[H_I,H_B]_n$ is defined by the iteration relation $[H_I,H_B]_n= [H_I,[H_I,H_B]_{n-1}]$ with $[H_I,H_B]_{n=1} \equiv [H_I,H_B]$, and  
$\left\langle O \right\rangle_\varrho \equiv \mathrm{tr}(\varrho O)$ [see Supplemental Material (SM)]. 
Equation~\eqref{n-d} characterizes the instantaneous heat transfer process of the model~\eqref{ham}. 
Especially, the sign and amplitude of the leading-order nonvanishing derivative indicates the instantaneous direction and speed of the heat flow, respectively. 
We note that the heat flow $\mathcal{Q}(t)$ is in general a non-analytic smooth function of time. 
Therefore, two flow functions $\mathcal{Q}_{1,2}(t)$ are not necessarily identical, even if their instantaneous characteristics at a special time $t_0$ are identical, that is, $d^n \mathcal{Q}_1/d t^n|_{t=t_0} = d^n \mathcal{Q}_2/d t^n|_{t=t_0}$ for any $n \geqslant 0$.

For revealing the effect of the intrasystem nonequilibrium, we assume that the system $A$ is composed of $N$ interacting subsystems $A_{1,2,\cdots,N}$~[See Fig.~\ref{mulitibody_carton}]. The Hamiltonian reads $H_A = H_{A0} + H_{AI}$, where $H_{A0}=\sum_{k=1}^{N} H_{A_k}$ with $H_{A_k}$ the Hamiltonian of the subsystem $A_k$, $H_{AI}$ describes the intra-subsystem interactions, and $H_{A0}$ and $H_{AI}$ do not need to commute.       

\textbf{Local-equilibrium states.} 
We employ the concept of the local-equilibrium (LE) states to characterize the initial hotness/coldness of the systems, in which ``temperatures'' are locally defined parameters. For the Hamiltonian in Eq.~\eqref{ham}, a LE state $\rho$ requires that the reduced density matrices of the subsystems $A_{k=1,2,\cdots,N}$ and $B$ take the form of Gibbs states~[See Fig.~\ref{mulitibody_carton}] 
\begin{equation} \label{les}
\rho_{A_k} = e^{-\beta_{A_k} H_{A_k}}/\mathcal{Z}_{A_k}, \quad \rho_B = e^{-\beta_B H_B}/\mathcal{Z}_B,
\end{equation}  
where $1/\beta_{u} \equiv T_{u}$ and $\mathcal{Z}_{u} = \tr \left[ \exp (-\beta_{u} H_{u}) \right]$ being the effective temperature and the partition function of the subsystem $u \in \{A_{1,2,\cdots,N}, B\}$, respectively, and we have taken the Boltzmann constant $k_B=1$. We denote the final state of the subsystem $u$ by $\rho_u'$. In practice, the LE states can be prepared by thermalizing the subsystems with independent baths. For later discussion, it is convenient to introduce the product states 
\begin{equation} \label{prods}
\rho_{A,0} \equiv \otimes_{k=1}^{N} \rho_{A_k}, \quad \rho_0 \equiv \rho_{A,0} \otimes \rho_B. 
\end{equation} 
The initial LE state $\rho$ is unnecessarily the product state $\rho_0$, which releases a crucial assumption for deducing entropy production \cite{esposito2010NJP} and Landauer's principle \cite{reeb2014NJP}. 
Without loss of generality, we assume that $A$ and $B$ systems are the cold and the hot, respectively, that is, $\beta_{A} > \beta_B$ where $ \beta_A \equiv \mathrm{min}\{\beta_{A_k}\}_{k=1}^N$.

\textbf{Classification of AHT mechanisms.}
We are able to express the heat transfer $\mathcal{Q}$ in terms of quantum relative entropy (QRE) and obtain the heat transfer equation        
\begin{equation}\label{AHT_multisystem}
\begin{aligned}
	(\beta_B-\beta_{A}) \mathcal{Q}
	= \Delta \mathcal{S} +\Delta \mathcal{I}_{AB}+\Delta \mathfrak{T}_A +\Delta \mathfrak{I}_A.
\end{aligned}
\end{equation}
The derivation of Eq.~\eqref{AHT_multisystem} and the connection between Eq.~\eqref{n-d} and Eq.~\eqref{AHT_multisystem} are presented in SM. 
In Eq.~\eqref{AHT_multisystem}, the left-hand side describes the effective entropy flux flowing from $A$ to $B$, and $(\beta_B-\beta_A) \mathcal{Q} < 0$ defines the AHT regime. The right-hand side characterizes various origins of the entropy flux. 
(I) $\Delta \mathcal{S} \equiv \sum_{k=1}^{N} S(\rho_{A_k}'||\rho_{A_k})+S(\rho'_B||\rho_B)$, in which each term describes the QRE between the final and initial states of an individual subsystem. Here $S(\varrho||\varsigma) \equiv -\tr (\varrho \ln \varsigma ) - S(\varrho)$ is the QRE between states $\varrho$ and $\varsigma$, with $S(\varrho) \equiv -\tr (\varrho \ln \varrho)$ the von-Neumann entropy of the state $\varrho$ \cite{nielsen2000}. 
(II) $\Delta \mathcal{I}_{AB} \equiv \mathcal{I}_{AB} (\rho') - \mathcal{I}_{AB} (\rho)$, in which $\mathcal{I}_{AB} (\rho) \equiv S(\rho|| \rho_0)$ is the multipartite mutual information (MI) among all the subsystems~\cite{watanabe1960IJRD,groisman2005PRA}, describes the variation of the subsystem correlation. 
(III) $\Delta \mathfrak{T}_{A} \equiv \sum _{k=1}^{N} (\beta _{A} -\beta _{A_k}) \Delta \langle H_{A_k} \rangle$ describes the entropy flux induced by the initial temperature inhomogeneity within the system $A$. 
(IV) $\Delta \mathfrak{I}_{A} \equiv \beta _{A} \Delta \langle H_{AI} \rangle$ characterizes the entropy flux generated by the interaction energy variation of the system $A$. 
We note that $S(\rho'_{A_k,B}||\rho_{A_k,B}) \geqslant 0$ by definition~\cite{nielsen2000}, while the signs of $\Delta \mathcal{I}_{AB}$, $\Delta \mathfrak{T}_A$, and $\Delta \mathfrak{I}_A$ are conditional. For example, $\Delta \mathcal{I}_{AB}<0$ requires initial-state correlations.    
In addition, we point out Eq. \eqref{AHT_multisystem} can be extended to the situations where the HTC is  broken (refer to SM for detailed discussions).

An immediate observation for Eq.~\eqref{AHT_multisystem} is that the heat transfer is normal, i.e., $(\beta_B-\beta_A) \mathcal{Q} \geqslant 0$, and the Clausius statement of the second law is recovered, when $A$ is interaction-free ($\Delta \mathfrak{I}_A =0$), and, initially, $A$ is temperature-homogenous ($\Delta \mathfrak{T}_A=0$) and quantum correlation is absent ($\Delta \mathcal{I}_{AB} =0$).

Most importantly, Eq.~\eqref{AHT_multisystem} implies that AHT occurs only if $\Delta \mathcal{I}_{AB}+\Delta \mathfrak{T}_A +\Delta \mathfrak{I}_A <0$, and, at least one of three mechanisms induces negative entropy fluxes. (I) \emph{Initial-state correlations} giving $\Delta \mathcal{I}_{AB}<0$. Since the evolution of the joint system is unitary, the MI variation $\Delta \mathcal{I}_{AB}$ can be further decomposed into the intra- and intersystem components: $\Delta \mathcal{I}_{AB}=\Delta \mathcal{I}_{A} + \Delta \mathcal{I}_{A:B}$. 
Here $\Delta \mathcal{I}_{A} \equiv \mathcal{I}_{A} (\rho_A') - \mathcal{I}_{A} (\rho_A)$, with $\mathcal{I}_{A} (\rho_A) \equiv S(\rho_A|| \rho_{A,0})$ the multipartite MI within $A$, gives the MI variation in $A$, and $\Delta \mathcal{I}_{A:B} \equiv \mathcal{I}_{A:B} (\rho') - \mathcal{I}_{A:B} (\rho)$, with $\mathcal{I}_{A:B} (\rho) \equiv S(\rho_A) + S(\rho_B) - S(\rho)$ the bipartite MI \cite{nielsen2000}, captures the variation of the total correlation between $A$ and $B$ \cite{modi2012RMP}. 
(II) \emph{Temperature inhomogeneity} giving $\Delta \mathfrak{T}_A <0$. (III) \emph{Intrasystem interactions} giving $\Delta \mathfrak{I}_A <0$. 
We note that $\Delta \mathcal{I}_{AB}+\Delta \mathfrak{T}_A +\Delta \mathfrak{I}_A <0$ is insufficient for AHT. For example, even if $\Delta \mathcal{I}_{A:B} <0$, the entropy production $\Delta \mathcal{I}_{A:B}+ S(\rho_B'||\rho_B)$ can still be positive due to large $S(\rho_B'||\rho_B)$, which likely produces a normal heat flow~\cite{esposito2010NJP, landi2021RMP, deffner2011PRL}.

The AHT dominated by only one of the mechanisms, e.g., $\Delta \mathcal{I}_{AB} <0$ and all the other entropy flux terms being non-negative, is of particular interest. The AHT induced by consuming the initial intersystem MI, i.e., $\Delta \mathcal{I}_{A:B}<0$, has been widely discussed in previous studies~\cite{PhysRevE.77.021110, jevtic2012PRL, jennings2010PRE, bera2017NC, micadei2019NC, medinagonzalez2020PRA, holdsworth2022PRA}; Consuming the intrasystem correlation, i.e., $\Delta \mathcal{I}_{A} <0$, can also give rise to AHT~\cite{latune2019PRR}. In this work, we give exhaustive examples of single-mechanism-dominated AHT in qubit systems.

\textbf{Heat transfer in multi-qubit systems.} We develop heat transfer models in multi-qubit systems under constraint of the HTC, and identify the AHT mechanisms in these models. Consider the system $X \in \{ A,B \}$ is composed of $N_X$ qubits. The interaction-free Hamiltonian reads 
\begin{equation} \label{tb-free}
H_{X0} = \sum_{i=1}^{N_X} H_{X_i}, \quad H_{X_i}=-\frac{1}{2}  \omega_{X_i} \sigma_{X_i}^z,
\end{equation} 
where $\omega_{X_i} \geqslant 0$ and $\sigma^{x,y,z}_{X_i}$ are the frequency and the Pauli operators of the qubit $X_i$, respectively. 
A qubit model with arbitrary Zeeman fields is identical to that in Eq.~\eqref{tb-free} up to local qubit rotations. The interaction Hamiltonian between $A$ and $B$ takes the general form 
\begin{equation}
H_I=H_{AB}^I+H_{AAB}^I+H_{ABB}^I+\cdots,
\end{equation} 
where $H^I_{AB}$, $H^I_{AAB(ABB)}$, and $\cdots$, denote two-body (2B), three-body (3B), and higher order many-body interactions, respectively. Here we focus on 2B and 3B interactions.  The detailed derivation of the Hamiltonians is presented in SM.  

(I) \emph{2B intersystem interactions.} Specifying the intersystem interactions of the 2B form 
$H^I_{AB}=\sum_{k,l}\sum_{a,b} c_{kl}^{ab} \sigma_{A_k}^a  \sigma_{B_l}^b$ and the intrasystem interactions of the general multi-body form 
$H_{XI}=\sum_{i,j,\cdots,k}  \sum_{a,b,\cdots,c} J_{X,ij \cdots k}^{ab\cdots c} \,  \sigma_{X_i}^a \sigma_{X_j}^b \cdots \sigma_{X_k}^c$, we find that the HTC leads to two independent commutation relations $[H_{A0} +H_{B0}, H^I_{AB}]=0$ and $[H_{XI}, H^I_{AB}]=0$. 
The former relation reduces the intersystem interactions to 
\begin{equation}\label{Twobody_interaction}
\begin{aligned}
	H_{AB}^I=\sum_{k,l} \left( c_{A_kB_l} \sigma^-_{A_k} \sigma^+_{B_l} + \textrm{h.c.} \right), 
	\end{aligned}
\end{equation}
where $\sigma_{X_i}^\pm \equiv \left(\sigma_{X_i}^x \pm i \sigma_{X_i}^y \right)/2$ and the coupling constant $c_{A_k B_l}$ is non-zero only if the two qubits $A_k$ and $B_l$ are of equal frequency $\omega_{A_k}=\omega_{B_l}$. The latter commutation relation should lead to constraints on $H_{AB}^I$ in Eq.~(\ref{Twobody_interaction}) and $H_{XI}$. However, we neglect $H_{XI}$ since we find it does not contribute to the heat transfer [$\Delta \mathfrak{I}_A=0$ in Eq.~\eqref{AHT_multisystem}]. Deriving the instantaneous heat transfer equations for the model defined by Eqs.~\eqref{tb-free} and \eqref{Twobody_interaction}, we obtain a no-go theorem:       

\begin{theorem}\label{theorem1}
In the presence of only 2B intersystem interactions, instantaneous AHT is impossible if all the qubit triplets $X_m X_{k (\neq m)} Y_l$, where $(X,Y) \in \{ (A,B), \, (B, A) \}$, have no initial quantum coherence
\bibnote{Quantum coherence among $n$ qubits ($n \geqslant 2$) is encoded in the off-diagonal elements of the $n$-qubit reduced density matrix represented in the basis of the eigenstates of the interaction-free Hamiltonian $H_{A0}+H_{B0}$ (non-degenerate) defined in Eq.~\eqref{tb-free}.}.
\end{theorem}
 
The proof of Theorem~\ref{theorem1} can be found in SM. 
We formally decompose the initial state $\rho$ into three components, $\rho= \otimes_{X,k} \rho_{X_k} + \mathcal{D} + \chi$, 
where $\rho_{X_k}$ is the Gibbs state of the qubit $X_k$ as defined in Eq.~\eqref{les}, and $\mathcal{D}$ (diagonal) and $\chi$ (with diagonal entries all zero) encode the classical correlation and quantum coherence among the qubits, respectively.            
Theorem \ref{theorem1} implies that, if intersystem interactions are limited to the 2B type, the initial classical correlation $\mathcal{D}$ alone ($\mathcal{\chi}=0$) cannot induce AHT.
Furthermore, not any initial quantum coherence can induce AHT. For example, an initial state with the coherence component 
$\chi = (\otimes_{X,k} \sigma_{X_k}^+ + \textrm{h.c.} )$ does not exhibit AHT. 
Lastly, Theorem~\ref{theorem1} imposes constraints on the entropy flux terms in Eq.~\eqref{AHT_multisystem}. If no initial coherence in all the triplets, 
we have $\Delta \mathfrak{T}_A < 0$ (any of subsystems of $A$ gaining energy) and $\Delta \mathfrak{T}_A > -\Delta \mathcal{S}-\Delta \mathcal{I}_{AB}$ ($B$ losing energy), which gives a lower bound for the MI, $\Delta \mathcal{I}_{AB} > -\Delta \mathcal{S}$.

In general, the quantum coherence among a triplet $X_mX_kY_l$ can be decomposed into doublet components and triplet collective components nonexistent in any doublet~\cite{radhakrishnan2016PRL,ma2017PRA}. 
In correspondence, we find that the intersystem coherence in doublets $X_mY_l$ and $X_kY_l$ only contributes to the heat transfer rate $d\mathcal{Q}/d t$, while the intrasystem coherence in doublet $X_mX_k$ as well as the triplet collective coherence contribute to the instantaneous heat transfer convexity $d^2\mathcal{Q}/dt^2$ (See SM). The AHT induced by intersystem doublet coherence has been realized in nuclear spin systems~\cite{micadei2019NC}. 
When this intersystem doublet coherence vanishes ($d\mathcal{Q}/d t=0$), the intrasystem doublet coherence can also induce AHT ($d^2\mathcal{Q}/d t^2 >0$), which might be interesting for experimental verification.

(II) \emph{3B intersystem interactions.} We further investigate the qubit systems in which the intersystem interactions are of the 3B form $H_I=H^I_{AAB}+H^I_{ABB}$ and the intrasystem interactions are of 2B form. The HTC leads to three independent commutation relations $[H_{A0}+H_{B0},H^I_{AAB}]=[H_{A0}+H_{B0},H^I_{ABB}]=[H_{AI}+H_{BI}, H_I]=0$, which impose constraints on the interaction parameters.  

The first commutator leads to three types of interactions depending on the frequencies of the qubits involved. For any three qubits $A_{k,l}$ and $B_m$, the interaction denoted by $H_{A_k,A_l,B_m}^I$ takes the forms as follow. 
For $\omega_u=\omega_v+\omega_w$ or $\omega_v=\omega_w$, where $(u,v,w) \in$ $\big\{ (A_k,A_l,B_m)$, $(A_l,A_k,B_m)$, $(B_m,A_k,A_l) \big\}$,  we obtain $H_{A_k,A_l,B_m}^I = c_{uvw} \sigma_{u}^+  \sigma_{v}^- \sigma_{w}^- + \textrm{h.c.}$ or $H_{A_k,A_l,B_m}^I = c_{uvw}' \sigma_u^z \, \sigma_v^- \sigma_w^+ + \mathrm{h.c.}$, respectively, where $c_{uvw}$ and $c_{uvw}'$ are arbitrary complex constants.
For unconstrained frequencies, we obtain $H_{A_k,A_l,B_m}^I = c_{klm}'' \sigma_{A_k}^z \sigma_{A_l}^z \sigma_{B_m}^z$, where $c_{klm}''$ is an arbitrary real constant. The second commutator gives $H^I_{ABB}$ of the similar form as $H_{AAB}^I$. The third commutator introduces further constraints to interaction parameters. We note that, unlike the 2B interaction case, the intersystem interaction terms $\sigma_{A_k}^z \sigma_{A_l}^z \sigma_{B_m}^z$ as well as the intrasystem interactions $H_{XI}$ cannot be omitted since they can contribute to heat transfer. We note that, exploiting 3B intersystem interactions, all the three types of AHT can be realized. In particular, AHT may occur even if the initial state is the product state $\rho_0$ in Eq.~\eqref{prods}. 
 
\textbf{AHT in three-qubit systems.} 
We study three-qubit system $\{A_{1,2}, B\}$ with a combination of 2B and 3B intersystem interactions. The derivation of the Hamiltonians and designing details of the AHT mechanism can be found in SM. 

\begin{figure}[t]
\centering
\subfigure{
\begin{minipage}[]{0.495\textwidth}
\begin{overpic}[width=1\textwidth]{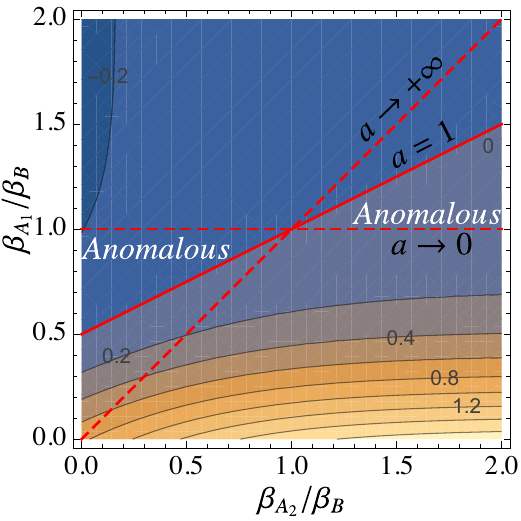}\label{triparitite_contour}
 \put(-2,94){\scriptsize{\textsf{(a)}}}
\end{overpic}
\end{minipage}
}
\hspace{-0.02\textwidth}
\subfigure{
\begin{minipage}[]{0.45\textwidth}	
\begin{overpic}[width=1\textwidth]{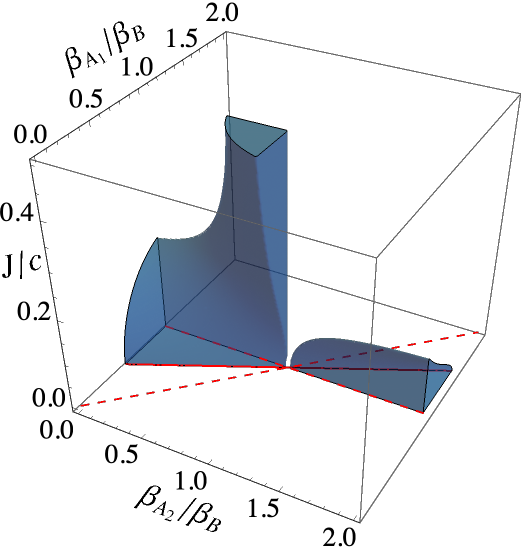}\label{triparitite_sustain}
 \put(0,96){\scriptsize{\textsf{(b)}}}
\end{overpic}
\end{minipage}
}	
\caption{
(a) Contour plot for the heat transfer convexity at the initial time [Eq.~\eqref{de2dt2_by_HIt}]. The AHT region lies between the red solid line ($a=1$), and the red dashed line ($a \rightarrow 0$). 
(b) AHT region (blue) when the 3B interaction (\ref{hiii}) is perturbed by a 2B interaction with strength $J$. Panel (a) corresponds to $J/c=0$. }
\label{triparitite_action}
\end{figure} 

(I) In Fig.~\ref{ThreeQubits_InnerTemper}, we show an example of temperature-inhomogeneity-dominant AHT. We take $H_{AI}=0$ and the intersystem 3B interaction takes the simple expression   
 \begin{equation} \label{hiii}
 	 	H_{A_1A_2B}^I=c \, \sigma_{A_1}^+ \sigma_{A_2}^- \sigma_{B}^- + \textrm{h.c.} \ ,
 \end{equation}
where $c$ is the coupling strength. We choose the initial state as the product state $\rho_0$ in Eq.~\eqref{prods}. Therefore, the MI variation $\Delta  \mathcal{I}_{AB} \geqslant 0$, and the intrasystem interaction contribution $\Delta \mathfrak{I}_A=0$. According to Eq. \eqref{AHT_multisystem}, when $-\Delta \mathfrak{T}_A > \Delta \mathcal{S}+\Delta \mathcal{I}_{A B}$, AHT occurs.

For this model the initial heat transfer rate vanishes $d\mathcal{Q}/dt=0$. 
Phase transition between the AHT and the normal heat transfer is shown in Fig. \ref{triparitite_contour}, which we can obtain by calculating the initial heat transfer convexity,   
\begin{equation}
\label{de2dt2_by_HIt}
\begin{aligned}
\frac{d^2\mathcal{Q}}{dt^2}
		=&\frac{2 |c|^2 \omega_{B}  \left[e^{ (\beta_{A_2}\omega_{A_2} + \beta_{B} \omega_{B} ) } - e^{ \beta_{A_1} \omega_{A_1} } \right] } {\left( 1+e^{\beta_{A_1} \omega_{A_1}  } \right) \left(1+e^{  \beta_{A_2} \omega_{A_2} } \right) \left(  1+e^{ \beta_{B} \omega_{B}  } \right) }, 
\end{aligned}
\end{equation} 
and further parametrizing  the qubit frequencies as $\omega_{A_1}=(1+a)\omega_{B}$ and $\omega_{A_2}=a \omega_{B}$ with $a \in (0, +\infty)$ (see SM). The phase boundary is defined by $d^2\mathcal{Q}/dt^2=0$, which gives $(1+a)\beta_{A_1}/\beta_{B}= 1+a \beta_{A_2}/\beta_{B}$ [the red lines in Fig.~\ref{triparitite_contour}].  
As $a$ varies from $0$ to $\infty$, the slope of the phase boundary varies from $0$ to $1$ correspondingly.  In the top-right region where $\beta_{A_2} > \beta_{A_1} > \beta_{B}$, for a fixed $\beta_{A_1}$, $\Delta \mathfrak{T}_A$ gets negative enough, such that $-\Delta \mathfrak{T}_A > \Delta \mathcal{S}+\Delta \mathcal{I}_{AB}(>0)$,  to realize AHT when $\beta_{A_2}$ is large enough to cross the phase boundary. 

As the 3B interaction (\ref{hiii}) is perturbed by weak 2B interactions, which breaks the HTC, the AHT is robust. For example, we take the 2B perturbation $H_b^I=J \sum_{u,v} \sum_{a} \sigma_u^a \sigma_v^a $ with a uniform coupling constant $J$. 
It can be proven that $H_b^I$ gives no correction to the initial heat transfer rate, while it gives a correction of the order of $J^2$ to the convexity (see SM). As shown in Fig. \ref{triparitite_sustain}, when $J/c=0$, we recover the phase diagram of the unperturbed system [Fig.~\ref{triparitite_contour}]. As $J/c$ increases, the region of AHT shrinks but remains finite.

(II) In Fig. \ref{ThreeQubits_InnerEnergy}, we show an example of intrasystem-interaction-dominated AHT. 
 We take the intersystem interaction $H_I = 0.005 \omega_0 \big[  \big(  2 i \sigma_{A_1}^- \sigma_{A_2}^+ + \mathrm{h.c.} \big) \sigma^y_{B}     
  +  \big( \sigma^z_{A_2} - \sigma^z_{A_1} \big) \sigma^x_{B} \big]$ being a combination of 2B and 3B types, and the intrasystem interaction $H_{AI}=  0.5  \omega_0 \big( \sigma_{A_1}^- \sigma_{A_2}^+ + \textrm{h.c.} \big)$. We choose the product state $\rho_0$ in Eq.~\eqref{prods} as the initial state with the qubits $A_{1,2}$ are of the same temperature $\beta_A = \beta_{A_{1,2}}$. Hence $\Delta \mathcal{I}_{AB} \geqslant 0$ and $\Delta \mathfrak{T}_A=0$ during the evolution. We observe that $ - \Delta \mathfrak{I}_A  > \Delta  \mathcal{S}+\Delta \mathcal{I}_{A B}   $ can induce AHT.

(III) In Fig.~\ref{ThreeQubits_InnerCorre}, we show an example of intrasystem-correlation-dominated AHT. The intrasystem interaction is set to be zero $H_{AI}=0$.
We prepare the initial state $\rho=(\rho_{A_1} \otimes \rho_{A_2} +\chi_A ) \otimes \rho_{B}$, where the coherence term $\chi_A=0.24 \left(\sigma_{A_1}^+ \otimes \sigma_{A_2}^- + \textrm{h.c.}\right)$ encodes quantum correlation within $A$ and $A_{1,2}$ share the same temperature $\beta_A = \beta_{A_{1,2}}$.  The joint system evolves under the 2B interaction $H_I=0.005 \omega_0 \left(\sigma_{A_1}^-\sigma_{B}^+ + \sigma_{A_2}^-\sigma_{B}^+  + \textrm{h.c.} \right)$. Obviously, we have $\Delta \mathfrak{T}_A=\Delta \mathfrak{I}_A=0$ and $\Delta \mathcal{I}_{A:B} \geqslant 0$ during the evolution.
Hence AHT can only be realized by consuming the correlation between the qubits $A_1$ and $A_2$, such that $- \Delta \mathcal{I}_{A} > \Delta \mathcal{S} + \Delta \mathcal{I}_{A:B}$, which is consistent with Theorem \ref{theorem1}. This mechanism has been discussed in Ref.~\onlinecite{latune2019PRR}.

{\bf Discussion.} 
We have shown that both intrasystem temperature inhomogeneity and intrasystem interactions are able to induce AHT in quantum systems, aside from initial inter- and intrasystem correlation. For qubit systems in which the interactions are limited to the 2B type, we have proven that the AHT occurs only if the initial states contain quantum coherence. Realizing AHT via the other mechanisms requires 3B interactions, which are usually weak and overwhelmed by 2B interactions in nature. However, 3B and multi-body interactions have been simulated in laboratory with various quantum techniques from trapped ions to superconducting qubits~\cite{maslennikov2019NC,buchler2007NP,zhao2017PRA,mezzacapo2014PRL,Chancellor2017}. It would be interesting to simulate the AHT phenomenon dominated by the mechanisms beyond the initial-state correlation.          

A quantum system exhibiting AHT is a potential platform for engineering heat pump, quantum absorption refrigerator,  and quantum autonomous thermal machines  \cite{linden2010PRL,levy2012PRL,du2018NJP,bhandari2021PRB,woods2023PRX}. A qubit heat pump driven entirely by quantum correlation as fuel has been proposed theoretically~\cite{holdsworth2022PRA}, and quantum refrigerators utilizing three thermal baths have been demonstrated~\cite{krishnan2024,huang2024PRL}. Moreover, thermodynamic computing using autonomous quantum thermal machines, where different operation regimes are tuned by temperature difference,  has been proposed \cite{lipka-bartosik2024SA}.  Our AHT classification offers an atlas for searching the heat engine cycles and optimizing the efficiency performance, in particular, by exploring the intrasystem degrees of freedom.

{\bf Acknowledgments.} 
We would like to thank Peng Zhao, Deepak Karki, Miklos Horvath, Thomas Lane, Fei Meng, Mingjing Zhao, and Yunxiang Liao for stimulating discussions. 
The work of T.M. is supported by NSFC Grant No.~11905100. 
The work of H.-Y.X. is supported by NSFC Grant No.~12074039. 
The work of J.-N.Z. is supported by NSFC Grant No.~92065205. 
The work of M.-H.Y. is supported by Natural Science Foundation of Guangdong Province Grant No.~2017B030308003, Key-Area Research and Development Program of Guangdong Province Grant No.~2018B030326001, Science, Technology, and Innovation Commission of Shenzhen Municipality Grant Nos.~JCYJ20170412152620376, JCYJ20170817105046702, and KYTDPT20181011104202253, NSFC Grant Nos.~11875160 and U1801661, Economy, Trade, and Information Commission of Shenzhen Municipality Grant No.~201901161512, and Guangdong Provincial Key Laboratory Grant No.~2019B121203002.
The work of H.-Y.X. was completed in the University of Oklahoma.




%


\onecolumngrid
\newpage

\renewcommand{\thetable}{S\arabic{table}}  
\renewcommand{\thefigure}{S\arabic{figure}} \setcounter{figure}{0}
\renewcommand{\theequation}{S\arabic{equation}}\setcounter{equation}{0}

\centerline{ \bf \large  Anomalous Heat Transfer in Nonequilibrium Quantum Systems}
\vspace{2pt}
\centerline{\bf \large SUPPLEMENTAL MATERIAL}

\onecolumngrid
\smallskip


\section{Derivation of Eq.~(2)}
From the Liouville-von Neumann equation $d\varrho/dt =i [\varrho,H]$, 
we obtain $(-i)^n d^{n}\varrho/dt^{n} = {_n}[ \varrho, H]$, where the operator ${_n}[ \varrho, H]$ is defined by the iteration relation ${_n}[ \varrho, H] = [{_{n-1}}[ \varrho, H],H]$ with ${_1}[ \varrho, H] \equiv [ \varrho, H]$. The $n_\mathrm{th}$ derivative of heat transfer reads 
\begin{equation} \label{dnE/dnt}
\begin{aligned}
(-i)^n\frac{d^{n}\mathcal{Q}}{dt^{n}} \equiv &\, (-i)^n\mathrm{tr} \left(\frac{d^{n} \varrho }{dt^{n}} H_B \right) \\ 
= &\, \mathrm{tr} \left( {_n}[ \varrho, H] H_B \right) 
= \mathrm{tr} \left( [{_{n-1}}[ \varrho, H],H] H_B \right) = \mathrm{tr} \left( {_{n-1}}[ \varrho, H] [H, H_B]_1 \right) = \mathrm{tr} \left( \varrho [H, H_B]_n \right),   
\end{aligned}
\end{equation} 
where we have applied the relation $\mathrm{tr} \left( {_{m}}[ \varrho, H] [H, H_B]_n \right) = \mathrm{tr} \left( {_{m-1}}[ \varrho, H] [H, H_B]_{n+1} \right)$ in the last equality. Since $[H,H_B] = [H_I,H_B]$, we obtain Eq.~(2).

\section{Derivation of Eq.~(5)}
We derive the heat transfer equation (5), exploiting the concept of quantum relative entropy of state $\varrho$ with respect to state $\varsigma$, i.e., $S(\varrho||\varsigma) \equiv -\tr (\varrho \ln \varsigma ) - S(\varrho)$, where $S(\varrho) \equiv -\varrho \ln \varrho$ is the von-Neumann entropy of the state $\varrho$. Especially, if the reference state $\varsigma$ is an equilibrium state, $S(\varrho||\varsigma)$ describes the ``distance'' of the state $\rho$ to the thermal equilibration. For a generic system $X$, described by Hamiltonian $H_X$ and evolving from the initial state $\rho_X$ to the final state $\rho_X'$, we define the relative entropy variation and the energy variation by
\begin{equation} \label{entropy-production}
\Sigma_{X} \equiv S(\rho'_X || \rho_X^\ast)-S(\rho_X||\rho_X^\ast), \quad \Delta \langle H_X \rangle \equiv \mathrm{tr}[(\rho_X'-\rho_X) H_X],
\end{equation}      
respectively, where the reference state $\rho_X^\ast$ is a Gibbs state $\rho_X^\ast \equiv e^{-\beta_{X}^\ast H_X}/\mathcal{Z}_{X}$ with $1/\beta_{X}^\ast$ the effective temperature and $\mathcal{Z}_X \equiv \tr e^{-\beta_X^\ast H_X}$ the partition function. We note that $-\Sigma_{X}$ is the entropy production if $\rho_X^\ast$ is the stationary state of $X$. 
Equation~(\ref{entropy-production}) leads to
\begin{equation} \label{entropy-production-2}
\beta_X^\ast  \Delta \langle H_X \rangle = \Delta S_X+\Sigma_{X},
\end{equation}
where $\Delta S_X = S(\rho_X')-S(\rho_X)$ denoting the entropy variation.  

For the heat transfer model in Eq.~(1), we define $\rho_{X}$ and $\rho_{X}'$ as the reduced density matrix of the system $X \in \{A,B\}$ for the initial state $\rho$ and the final state $\rho'$, respectively. Applying Eq.~\eqref{entropy-production-2} to $A$ and $B$ systems, we obtain the relation 
\begin{equation} \label{arrowoftimenoneq}
(\beta_B^\ast-\beta_A^\ast) \mathcal{Q} = \Delta \mathcal{I}_{A:B} + \Sigma_{A} + \Sigma_{B},
\end{equation}
where $\Delta \mathcal{I}_{A:B} \equiv \mathcal{I}_{A:B}(\rho')-\mathcal{I}_{A:B}(\rho)$, with $\mathcal{I}_{A:B}(\rho) \equiv S(\rho_A)+S(\rho_B)-S(\rho)$ the bipartite mutual information, captures the variation of the total correlation between $A$ and $B$, and we have used the relation $S(\rho')=S(\rho)$ since the joint system is isolated. Note that the reference temperatures $\beta_{A,B}^\ast$ are arbitrary and need to be specified according to physical circumstances. 

As $A$ is composed of $N$ subsystems $A_{1,2,\cdots,N}$ and initially in local equilibrium [Eq.~(3)], the relative entropy variation of $A$ can be further decomposed. From Eq.~\eqref{entropy-production-2} we obtain
\begin{equation}
\begin{aligned}\label{SigmaA}
\Sigma _{A}= & -\Delta S_{A} + \beta_A^\ast \sum _{k=1}^{N} \Delta \langle H_{A_{k}} \rangle + \beta_A^\ast \Delta \langle H_{AI} \rangle\\
= & \sum _{k=1}^{N} \big( \beta _{A_{k}} \Delta \langle H_{A_{k}} \rangle - \Delta S_{A_{k}} \big) +\left( \sum_{k=1}^{N} \Delta S_{A_{k}} -\Delta S_{A}\right)+\sum _{k=1}^{N}( \beta_A^\ast -\beta _{A_{k}}) \Delta \langle H_{A_{k}} \rangle+\beta_A^\ast \Delta \langle H_{AI} \rangle  \\
= & \sum _{k=1}^{N} S( \rho '_{A_{k}} ||\rho _{A_{k}}) +\Delta \mathcal{I}_{A} +\Delta \mathfrak{T}_{A} +\Delta \mathfrak{I}_{A}.
\end{aligned}
\end{equation}  
In the last equality of Eq.~\eqref{SigmaA}, we have defined $\Delta \mathcal{I}_{A} \equiv \mathcal{I}_{A} (\rho_A') - \mathcal{I}_{A} (\rho_A) $, with $\mathcal{I}_{A}(\varrho_A) \equiv S(\varrho_A||\rho_{A,0}) = \sum_{k=1}^N S(\rho_{A_k}) - S(\varrho_A)$ the relative entropy among the $N$ subsystems, $\Delta \mathfrak{T}_{A} \equiv \sum _{k=1}^{N} (\beta_{A}^\ast -\beta _{A_k} ) \Delta\langle H_{A_k}\rangle$, and $\Delta \mathfrak{I}_{A} \equiv \beta _{A}^\ast \Delta \langle H_{AI} \rangle$. 
For the system $B$, we identify the reference state $\rho_B^\ast$ to the Gibbs state $\rho_B$ in Eq.~(3) and obtain 
\begin{equation} \label{SigmaB}
\Sigma_B=S(\rho_B'||\rho_B), \quad \beta_B^\ast = \beta_B. 
\end{equation} 
Assuming that $A$ and $B$ are the cold and the hot, respectively, i.e., $1/\beta_A < 1/\beta_B$ with $1/\beta_A = \mathrm{max}\{ 1/\beta_{A_k}\}$, we take the reference temperature $\beta_A^\ast=\beta_A$. Combining Eqs.~\eqref{arrowoftimenoneq}-\eqref{SigmaB}, we obtain the heat transfer equation (5). 

Via the Liouville-von Neumann equation, we obtain the differential equations for the entropy-flux terms on the right-hand side of Eq.~(5),
\begin{subequations} \label{decomposition2}
\begin{align}	
& \dot{S}(\varrho_{A_k}(t)||\rho_{A_k})=\beta_{A_k} \dot E_{A_k}(t) - \dot S_{A_k}(t),
  \quad \dot S(\varrho_B(t)||\rho_B) =  \beta_{B} \dot E_{B}(t) - \dot S_{B}(t), \nonumber \\
& \Delta \dot{\mathcal{I}}_{A} = \sum_{k=1}^N \dot S_{A_k}(t) - \dot S_{A}(t), \quad 
  \Delta \dot{\mathcal{I}}_{A:B} = \dot S_{A} (t) + \dot S_{B} (t), \quad 
  \Delta \dot{\mathfrak{T}}_A = \sum_{k=1}^N \left( \beta_{A} - \beta_{A_{k}} \right) \dot E_{A_k}(t), \quad
  \Delta \dot{\mathfrak{I}}_A = \beta_A \dot E_{AI}(t), 
\end{align}
and, $\Delta \dot{\mathcal{I}}_{AB} \equiv \Delta \dot{\mathcal{I}}_{A} + \Delta \dot{\mathcal{I}}_{A:B} = \sum_{k=1}^N \dot S_{A_k}(t) + \dot S_{B} (t)$, where we have introduced the dot notation for time derivative and
\begin{align}
  \dot E_{X}(t) = i \big\langle [H, H_{X}] \big\rangle_{\varrho(t)}, \quad
  \dot S_{X}(t) =  -i \big\langle [ H, \ln \varrho_{X}(t)] \big\rangle_{\varrho(t)}, \quad
  \dot E_{AI}(t) =  i \big\langle [H, H_{AI}] \big\rangle_{\varrho(t)}, \label{SE-dt}
\end{align} 
\end{subequations} 
with $\varrho(t)$ being the instantaneous state in Eq.~(2), $X$ denoting any of the (sub)systems, and $\varrho_{X}(t) \equiv \mathrm{tr}_{\setminus X} \varrho(t)$ being the reduced density matrix of $X$. Combining Eqs.~(5) and \eqref{decomposition2} one recovers Eq.~(2). We note that for $n_\mathrm{th}$ ($n \ge 2$) time derivative $d^n E_{X(AI)}(t)/dt^n = i^n \langle [H,H_{X(AI)}]_n \rangle_\varrho$,  while $d^n S_{X}(t)/dt^n$ has no simple expressions without further assumptions since the equations of motion of $\varrho_X(t)$ are involved.

\section{Heat transfer conditions (HTC) in qubit systems}
 
(I) \emph{Two-body intersystem interactions.} 
For qubit systems with the two-body (multi-body) intersystem (intrasystem) interactions $H_{AB}^I$ ($H_{XI}$) defined in the main text, the heat transfer condition gives 
\begin{equation} \label{htc-1}
[H_{A0}+H_{B0},H_{AB}^I]+[H_{AI},H_{AB}^I]+[H_{BI},H_{AB}^I]=0.
\end{equation}
We obtain the commutators
\begin{equation}\label{commutator_form}
\begin{aligned}	
[H_{A0} + H_{B0}, H_{AB}^I] \sim \sigma_A \otimes \sigma_B, \quad
[H_{AI}, H_{AB}^I] \sim \sigma_A \otimes \cdots  \sigma_A \otimes \sigma_B, \quad 
[H_{BI}, H_{AB}^I] \sim \sigma_A \otimes \sigma_B \otimes \cdots  \sigma_B,
\end{aligned}
\end{equation}
where ``$\sigma_X \otimes \cdots \sigma_Y$'' represents a linear combination of the tensor products $\{\sigma_{X_i}^a \cdots \sigma_{Y_j}^b\}$. Note that the commutators in Eq.~\eqref{commutator_form} are linearly independent. Combining Eqs.~\eqref{htc-1} and \eqref{commutator_form}, we readily have 
\begin{equation} \label{2b-cons}
[H_{A0}+H_{B0},H_{AB}^I]=[H_{XI},H_{AB}^I]=0.
\end{equation}

(II) \emph{Three-body intersystem interactions.} For three-body intersystem interactions $H_I=H^I_{AAB} + H^I_{ABB}$, the heat transfer condition leads to 
\begin{equation} \label{3b-htc}
[H_{A0}+H_{B0},H^I_{AAB}]+[H_{A0}+H_{B0},H^I_{ABB}]+[H_{AI}+H_{BI},H_I] =0.
\end{equation}
Assuming that the intrasystem interactions are of two-body type $H_{XI}=\sum_{i,j} \sum_{a,b} J_{X,ij}^{ab} \sigma_{X_i}^a \sigma_{X_j}^b$ and the three-body intersystem interactions take the form $H^I_{XYZ} \equiv \sum_{k,l,m} \sum_{a,b,c} c_{X_k Y_l Z_m}^{abc} \sigma_{X_k}^a \sigma_{Y_l}^b \sigma_{Z_m}^c$ where $XYZ \in \{ AAB, ABB \}$, we obtain the commutators
\begin{equation} \label{3b-c}
\begin{aligned}
& [H_{A0} +H_{B0}, H^I_{AAB}] \sim \sigma_A \otimes \sigma_A \otimes \sigma_B, \quad
  [H_{A0} +H_{B0}, H^I_{ABB}] \sim \sigma_A \otimes \sigma_B \otimes \sigma_B, \\
& [H_{AI},H^I_{AAB}] \sim \sigma_A \otimes \sigma_A \otimes \sigma_A \otimes \sigma_B \,\, \textrm{or} \,\, \sigma_A \otimes \sigma_B, \quad
  [H_{AI},H^I_{ABB}] \sim \sigma_A \otimes \sigma_A \otimes \sigma_B \otimes \sigma_B, \\
& [H_{BI},H^I_{AAB}] \sim \sigma_A \otimes \sigma_A \otimes \sigma_B  \otimes \sigma_B, \quad
  [H_{BI},H^I_{ABB}] \sim \sigma_A \otimes \sigma_B \otimes \sigma_B \otimes \sigma_B ~ \textrm{or} ~ \sigma_A \otimes \sigma_B.
\end{aligned}
\end{equation}
From Eq.~\eqref{3b-c} we observe that the commutators in Eq.~\eqref{3b-htc} are linearly independent, and, therefore, 
\begin{equation} \label{3b-cons}
[H_{A0}+H_{B0},H^I_{AAB}]=[H_{A0}+H_{B0},H^I_{ABB}]=[H_{AI}+H_{BI},H_I]=0.
\end{equation} 
The heat transfer conditions in Eqs.~\eqref{2b-cons} and \eqref{3b-cons} leads to constrain the parameters of the Hamiltonians.

First, both Eqs.~\eqref{2b-cons} and \eqref{3b-cons} involve the constraints of the type $[H_{A0}+H_{B0},H_I]=0$. We can find the solution of $[H_{A0}+H_{B0},H_I]=0$ for an arbitrary type of $H_I$. In general, $H_I$ can be formally written as  
\begin{equation}\label{multi-body_HI}
H_{I} =\sum _ {S_+,S_-,S_z} c_{S_{+} , S_{-} , S_{z}} \Xi_{{S_{+} , S_{-} , S_{z}}}, \quad \Xi_{{S_{+} , S_{-} , S_{z}}} \equiv \left( \prod_{u\in S_{+}} \sigma_{u}^{+}\right) \left( \prod_{v\in S_{-}} \sigma_{v}^{-}\right) \left( \prod_{w\in S_{z}} \sigma _{w}^{z}\right),
\end{equation}
where $S_a$ with $a \in \{\pm,z\}$ denotes a set of the indices of $\sigma^a$-type qubits, and $c_{S_{+},S_{-},S_{z}}^\ast=c_{S_-,S_+,S_z}$ due to Hermiticity. 
$\Xi_{{S_{+} , S_{-} , S_{z}}}$ represents a multi-body interaction of $(S_{+},S_{-},S_{z})$ type. We can easily calculate the commutator  
\begin{equation}\label{multi-body_commu}
\begin{aligned}
 \left[ H_{A0} + H_{B0} ,  H_{I} \right] 
 = -\sum _{S_{+},S_{-},S_{z}}  \Xi_{{S_{+} , S_{-} , S_{z}}}  c_{S_+,S_-,S_z}  \sum_{\sigma=\pm} \sigma 
 \left( \sum_{A_k \in S_\sigma} \omega _{A_{k}} + \sum_{B_k \in S_\sigma} \omega _{B_k} \right).
\end{aligned}
\end{equation} 
Hence $[ H_{A0} + H_{B0} ,  H_{I} ]=0$ leads to the constraint
\begin{equation}\label{frequency_matching}
c_{S_+,S_-,S_z}  =0, \quad \mathrm{or} \quad \sum_{A_k \in S_+} \omega _{A_{k}} + \sum_{B_k \in S_+} \omega _{B_k}  = \sum_{A_k \in S_-} \omega _{A_{k}} + \sum_{B_k \in S_-} \omega _{B_k},
\end{equation}
for any configuration $(S_{+},S_{-},S_{z})$, due to the linear independency of the operators $\Xi_{{S_{+} , S_{-} , S_{z}}}$. We apply Eq.~\eqref{frequency_matching} to the cases of two- and three-body interactions.


(I) \emph{Two-body intersystem interaction $H_{AB}^I$.} $[H_{A0}+H_{B0},H_{AB}^I]=0$ leads to Eq.~(8). We note that we have neglected the interaction terms $\sum_{k,l} c_{kl}^{zz}\sigma_{A_k}^z \sigma_{B_l}^z$, since these terms do not contribute to the instantaneous heat transfer. This can be readily proved by Eq.~(2). The commutator $[H_{XI},H_{AB}^I]=0$ in Eq.~\eqref{2b-cons} should introduce further constraints to interaction parameters. Nevertheless, from Eq.~(2) one can prove that $H_{XI}$ gives no contribution to the instantaneous heat transfer. Therefore, we have neglected the intrasystem interactions, $H_{XI}=0$.

(II) \emph{Three-body intersystem interaction $H_{AAB}^I$ and $H_{ABB}^I$.} From Eq.~\eqref{frequency_matching}, we find that the commutator $[H_{A0} + H_{B0}, H_{AAB}^{I}]=0$ in Eq.~\eqref{3b-cons} leads to the expression of $H_{ABB}^I$ presented in the main text. $[H_{A0}+H_{B0},H^I_{ABB}]=0$ gives $H^I_{ABB}$ of the similar form as $H_{AAB}^I$. In addition, the commutator $[H_{AI}+H_{BI},H_I]=0$ in Eq.~\eqref{3b-cons} introduces further constraints to interaction parameters.

\section{Proof and discussion of Theorem 1}
For the qubit Hamiltonian given by Eqs.~(6) and (8) and $H_{XI}=0$, we obtain 
\begin{equation}
\begin{aligned} \label{commutation_1order}
[H_{AB}^I,H_{B0}] = -\sum_{k,l} \omega_{B_l} \left( c_{A_kB_l}  \sigma_{A_k}^- \sigma^+_{B_l} - \mathrm{h.c.} \right). 
\end{aligned}
\end{equation}  
Therefore, by Eq.~(2) we conclude that the heat transfer rate $d \mathcal{Q}/d t$ vanishes if there is no intersystem coherence in any doublet $A_kB_l$, that is, the reduced density matrix $\varrho_{A_kB_l} \equiv \tr_{\setminus A_kB_l} \varrho$ is diagonal. 

From Eq.~(2) we obtain the heat transfer convexity 
\begin{equation}\label{second_derivative}
\begin{aligned}
 \frac{d^2 \mathcal{Q}}{d t^2}
= \sum_{k\neq m,l\neq n}   \frac{d^2 \mathcal{Q}_{mn,kl}}{d t^2} + \sum_{k\neq m,l}  \frac{d^2 \mathcal{Q}_{ml,kl}}{d t^2} +\sum_{k,l\neq n}  \frac{d^2 \mathcal{Q}_{kn,kl}}{d t^2} + \sum_{k,l}  \frac{d^2 \mathcal{Q}_{kl,kl}}{d t^2} \ ,
\end{aligned}
\end{equation}
where $d^2 \mathcal{Q}_{mn,kl}/d t^2  = -\left \langle  \left[ H_{A_mB_n}^I,\left[ H_{A_kB_l}^I, H_{B0} \right] \right] \right \rangle_{\varrho}$. We evaluate the commutators
\begin{equation} \label{2comm_Hij}
\begin{aligned}
& \left. \left[ H_{A_mB_n}^I,\left[ H_{A_kB_l}^I, H_{B0} \right] \right] \right|_{k \neq m}
= -\delta_{nl} \omega_{B_l} \left( c_{A_m B_l} c_{A_k B_l}^\ast  \sigma^-_{A_m} \sigma^+_{A_k}  + \mathrm{h.c.} \right) \sigma_{B_l}^z, \\
& \left[ H_{A_k B_l}^I,\left[ H_{A_kB_l}^I, H_{B0} \right] \right] = |c_{A_kB_l}|^2 \omega_{B_l} \left( \sigma_{A_k}^z - \sigma_{B_l}^z \right).
\end{aligned}
\end{equation}
The first term in Eq.~\eqref{second_derivative} obviously vanishes. 
The second term in Eq.~\eqref{second_derivative} vanishes if there is no coherence among any triplet $A_m A_k B_l$ (the triplet state is diagonal), since all the diagonal elements of the corresponding commutators in Eq.~(\ref{2comm_Hij}) are zero. Similarly, the third term in Eq.~\eqref{second_derivative} vanishes if there is no coherence among any triplet $A_kB_nB_l$. The last term in Eq.~\eqref{second_derivative} reads
\begin{equation}\label{Q_Hjl}
\begin{aligned}
 \frac{d^2 \mathcal{Q}_{kl,kl}}{d t^2}  
&= |c_{A_kB_l}|^2 \omega_{B_l}
	\Big [  \tr \left( \varrho_{B_l}    \sigma_{B_l}^z \right) - \tr \left( \varrho_{A_k}  \sigma_{A_k}^z \right)  \Big ] \ , 
\end{aligned}	
\end{equation}
where $\varrho_{X_k} = \tr_{ \setminus {X_k}}  \varrho $ is the reduced density matrix of $X_k$. Considering that all the qubits that initially in local equilibrium, $\rho_{X_k}  = e^{-\beta_{X_k} H_{X_k}}/\mathcal{Z}_{X_k}$, we find 
\begin{equation}
(\beta_{B_l} -	\beta_{A_k} )  \left. d^2 \mathcal{Q}_{kl,kl}/d t^2 \right| _{t=0} \geqslant 0,
\end{equation} 
with the equality holds only if $\beta_{A_k}=\beta_{B_l}$. Therefore, AHT is prohibited. 
Similarly one could obtain 
\begin{equation}
\left. d E_{A_k}/d t \right|_{t=0} = 0, \quad \left. d^2 E_{A_k}/d t^2 \right|_{t=0}  >   0,
\end{equation}
if there is no coherence among any triplet $A_kB_nB_l$ or $A_m A_k B_l$ (assuming $\beta_{A_k} > \beta_{B_l}$).

Both intra- and intersystem coherence can give rise to the second-order AHT. The initial state of the triplet can be formally written as 
\begin{equation}\label{AmAkBl_decomposition}
\begin{aligned}
\rho_{A_{m}A_{k}B_{l}} = \rho_{A_{m} A_{k}} \otimes \rho _{B_{l}} +  \sum_{a=x}^z \chi_{A}^a \otimes \sigma_{B_{l}}^a,
\end{aligned}
\end{equation} 
where $\tr(\chi_A^{x,y,z})=0$, the terms $\chi_{A}^a \otimes \sigma_{B_{l}}^a$ encode intersystem correlation, and $\rho_{B_l}$ is the Gibbs state of $B_l$ with temperature $1/\beta_{B_l}$. We obtain 
\begin{equation}
\begin{aligned}\label{d2Qdt2_AmAkBl_decomposition}
 \frac{d^2 \mathcal{Q}_{ml,kl}}{d t^2} 
= \omega_{B_l} \left[ \tanh \left( \frac{\beta_{B_l} \omega_{B_l}}{2} \right) \left\langle  c_{A_mB_n} c_{A_kB_l}^\ast \sigma^-_{A_m} \sigma^+_{A_k}   + \textrm{h.c.} \right\rangle_{\rho_{A_mA_k}} + 2 \left\langle c_{A_mB_n} c_{A_kB_l}^*  \sigma^-_{A_m} \sigma^+_{A_k}  + \textrm{h.c.} \right\rangle_{\chi_A^z} \right].
\end{aligned}	
\end{equation}
The first term represents the intrasystem coherence in $A$, and the second term represents the intersystem correlation, where only the collective coherence terms $\sigma_{A_m}^\pm \sigma_{A_k}^\mp \sigma_{B_l}^z$ can give finite contributions.

\section{Heat transfer in three-qubit systems}
For three qubit system $\{A_{1,2},B\}$ with a combination of two-body and three-body intersystem interactions $H_I=H_{AB}^I+H_{AAB}^I$, the heat transfer condition gives
\begin{equation} \label{3q-htc}
[H_{AAB}^I, H_{A0}+H_{B0}]+ [H_{AB}^I, H_{AI}] + [H_{AAB}^I, H_{AI}] + [H_{AB}^I, H_{A0}+H_{B0}] =0.
\end{equation}
In general the interaction Hamiltonians take the expressions 
$H_{AAB}^I=\sum_{a,b,c} c^{abc}  \sigma_{A_1}^a \sigma_{A_2}^b \sigma_{B}^c$, $H_{AB}^I=\sum_{i}\sum_{a,b} c_i^{ab} \, \sigma_{A_i}^a \sigma_B^b$, and 
$H_{AI}=\sum_{a,b} J^{ab} \sigma_{A_1}^a \sigma_{A_2}^b$. We obtain the commutators,
\begin{equation}
	\begin{aligned}
& [H_{AAB}^I, H_{A0} + H_{B0}] + [H_{AB}^I,  H_{AI}] \sim \sigma_{A_1} \otimes \sigma_{A_2} \otimes \sigma_{B}, \quad
  [H_{AAB}^I, H_{AI}] + [H_{AB}^I,  H_{A0} + H_{B0}] \sim \sigma_{A}   \otimes \sigma_{B}.
	\end{aligned}
\end{equation}
Therefore, Eq.~\eqref{3q-htc} leads to
\begin{equation} \label{3q-cons}
[H_{AAB}^I, H_{A0} + H_{B0}] + [H_{AB}^I,  H_{AI}] = [H_{AAB}^I, H_{AI}] + [H_{AB}^I,  H_{A0} + H_{B0}] =0.
\end{equation}	

We consider a special parametrization. We take the frequencies and the intrasystem interactions of the forms 
\begin{equation} \label{2B3BHAI}
\omega_{A_1}/r_1 = \omega_{A_2}=\omega_{B} \equiv \omega_0, \quad 
H_{AI}= \omega  \left( r_2 \sigma_{A_1}^+ \sigma_{A_2}^- + \mathrm{h.c.} \right),
\end{equation}
where $r_1 >0$ and $r_2$ is a complex number. We find that, for $r_2 = \pm \frac{1}{2} \sqrt{r_1 (2 - r_1) }$, the commutators in Eq.~\eqref{3q-cons} gives
\begin{equation} \label{23bodymixed_HI}
\begin{aligned}
& c^{xxx}= c^{yyx}=(1-r_1)  c^{xyy}, \quad 
  c^{xxy}= c^{yyy}=-(1-r_1) c^{xyx}, \quad
  c^{yxx}=-c^{xyx}, \quad 
  c^{yxy}=-c^{xyy}, \\ 
& c^{zx}_{1} = - 2 r_2  c^{xyy}, \quad 
  c^{zy}_{1} =   2 r_2  c^{xyx},    \quad
  c^{zx}_{2} =   2 r_2  c^{xyy}    \quad
  c^{zy}_{2}= -  2 r_2  c^{xyx},
\end{aligned}
\end{equation}
and all the other interaction parameters vanish.  From Eq.~(\ref{23bodymixed_HI}), we obtain the interactions
\begin{equation}\label{HI_23mixed}
\begin{aligned}
H_I = H_{AAB}^I + H_{AB}^I
    = & \, 2 i \left( \sigma_{A_1}^- \sigma_{A_2}^+ - \sigma_{A_1}^+ \sigma_{A_2}^- \right) \left(c^{xyy} \sigma^y_{B} + c^{xyx} \sigma^x_{B} \right) \\  
      & \, + \left[ 2 (1-r_1) \left( \sigma_{A_1}^- \sigma_{A_2}^+ + \sigma_{A_1}^+ \sigma_{A_2}^- \right)  + 2  r_2 \left(\sigma^z_{A_2} - \sigma^z_{A_1} \right) \right]  \left(c^{xyy} \sigma^x_{B}- c^{xyx} \sigma^y_{B} \right)  \ ,
\end{aligned}
\end{equation}
where the parameters $r_1 \in (0, 2]$, $r_2 = \pm \frac{1}{2} \sqrt{r_1 (2 - r_1) } \in [-1/2, 1/2] $, and $c^{xyy,xyx}$ are real. 
For $r_1=1$ and $r_2=1/2$, Eq.~(\ref{HI_23mixed}) reduces to the intersystem interaction Hamiltonian in Fig. 1(c).

\smallskip

For the model described by $\omega_{A_1}=\omega_{A_2}+\omega_{B}$, $H_{AI}=0$, and Eq.~(9), by Eq.~(2) we obtain the heat transfer rate and convexity
\begin{equation} \label{three-body_dQ/dt}
\frac{d \mathcal{Q}}{d t} =i  \omega_{B} \left\langle c  \sigma_{A_1}^+ \sigma_{A_2}^- \sigma_{B}^-  -   \mathrm{h.c.} \right\rangle_{\rho}, \quad
\frac{d^2 \mathcal{Q}}{d t^2} 
 = - \frac{1}{2} \omega_{B} |c|^2  \left\langle \sigma_{A_1}^z - \sigma_{A_2}^z -\sigma_{B}^z + \sigma_{A_1}^z  \sigma_{A_2}^z \sigma_{B}^z \right\rangle_{\rho}. 
\end{equation}
For a coherence-free initial state, i.e., diagonal $\rho$ in Eq.~(\ref{three-body_dQ/dt}), the initial heat transfer rate vanishes. 
When the initial state is the product state $\rho_0 =\rho_{A_1} \otimes \rho_{A_2} \otimes \rho_{B}$ defined in Eq.~(4), we obtain the initial heat transfer convexity in Eq.~(10). In addition, at the phase boundary $\beta_{B}=\beta_{A_1} +a(\beta_{A_1}-\beta_{A_2})$, we find that $d{\varrho}/dt|_{t=0}=0$.

The heat transfer condition $[H_I,H]=0$ is broken by adding the two-body interactions $H^I_b=J \sum_a\sum_{u,v} \sigma_u^a \sigma_v^a$, where $J$ is a real constant. For the product state $\rho_0$, it is obvious that the initial heat transfer rate vanishes. 
The heat convexity reads 
\begin{equation}\label{stability_tripartite_heat_transfer}
\begin{aligned}
		\frac{d^2 \mathcal{Q}}{dt^2}  = & \frac{4 |c|^2 \omega_{B}  \left(e^{2 \omega_{B} (a\beta_{A_2} + \beta_{B})} -e^{2(1+a)\omega_{B} \beta_{A_1}} \right) } {\left( 1+e^{2(1+a)\omega_{B} \beta_{A_1} } \right) \left(1+e^{2 a \omega_{B} \beta_{A_2}} \right) \left(  1+e^{2 \omega_{B} \beta_{B} } \right) } \\
		&- \frac{ 16 J^2 \omega_{B} }{1+e^{2\omega_{B}  \beta_{B}}} \left( \frac{e^{2 (1+a) \omega_{B} \beta_{A_1}} -e^{2 \omega_{B}  \beta_{B}}}{1+e^{2 (1+a) \omega_{B}  \beta_{A_1}}} + \frac{e^{2 a \omega_{B} \beta_{A_2}} -e^{2 \omega_{B} \beta_{B}}}{1+e^{2 a \omega_{B}  \beta_{A_2}}} \right) \ . 
\end{aligned}
\end{equation}
On the right side of Eq. (\ref{stability_tripartite_heat_transfer}), the first term is identical to Eq.~(10) and the second term is induced by  $H_b^I$. Therefore, 
the AHT region survives for sufficiently small $J$.

\section{Construction of AHT Hamiltonians}


(I) \textbf{Temperature inhomogeneity.} Our aim is to  design an intrasystem temperature inhomogeneity induced AHT without any initial correlations and intrasystem interaction. 
According to Eq. (5) in the main text, here due to $\Delta \mathcal{I}_{AB} \geqslant 0$ and $\Delta \mathfrak{I}_A = 0$,  we choose the product state,
\begin{equation}\label{productstate}
	\rho_{A_1} \otimes \rho_{A_2} \otimes  \rho_B  , 
\end{equation}
as the initial state, and set the intrasystem interaction to be zero, i.e., $H_{AI}=0$. We can parameterize the temperatures as 
\begin{equation}\label{parameterize-TA1TA2}
	T_{A_1}= e \, T_{B}, \ \ \  T_{A_2}=f \, T_B, 
\end{equation}
where $e,f \in (0,1) $, and assuming $e > f$ to ensure the intrasystem  temperature inhomogeneity. 
Considering a combination of 2B and 3B interaction Hamiltonian $H_I=H^I_{A_1B}+H^I_{A_1A_2B}$,  we find  $H^I_{A_1B}$ and $H^I_{A_1A_2B}$ is of the form of Eq. (8) and Eq. (9), respectively.  This is observed by Eq. \eqref{3q-cons}, $H_{AI}=0$, and Eqs.  \eqref{multi-body_HI}-\eqref{frequency_matching}.  
However Theorem 1 has already pointed out that AHT is impossible with 2B interaction, hence we should consider the heat transfer by the key 3B interaction $H^I_{A_1A_2B}$, i.e., Eq. (9):
\begin{equation} \label{HI-(I)}
 	H^I_{A_1A_2B}= c \sigma_{A_1}^{+} \sigma_{A_2}^{-} \sigma_B^{-}+\text {h.c. }
\end{equation}
For this Hamiltonian, according to Eq. \eqref{frequency_matching}, we parameterize the frequencies of qubits as   
\begin{equation}
	\omega_{A_1}=(1+a)\omega_0, \ \ \ \omega_{A_2}=a \omega_0, \ \ \ \ \omega_B \equiv \omega_0,  
\end{equation}
where $a\in (0, +\infty)$. 

\begin{figure}[htb]
\centering
\subfigure{
\begin{minipage}[]{0.49\textwidth}
\begin{overpic}[width=1\textwidth]{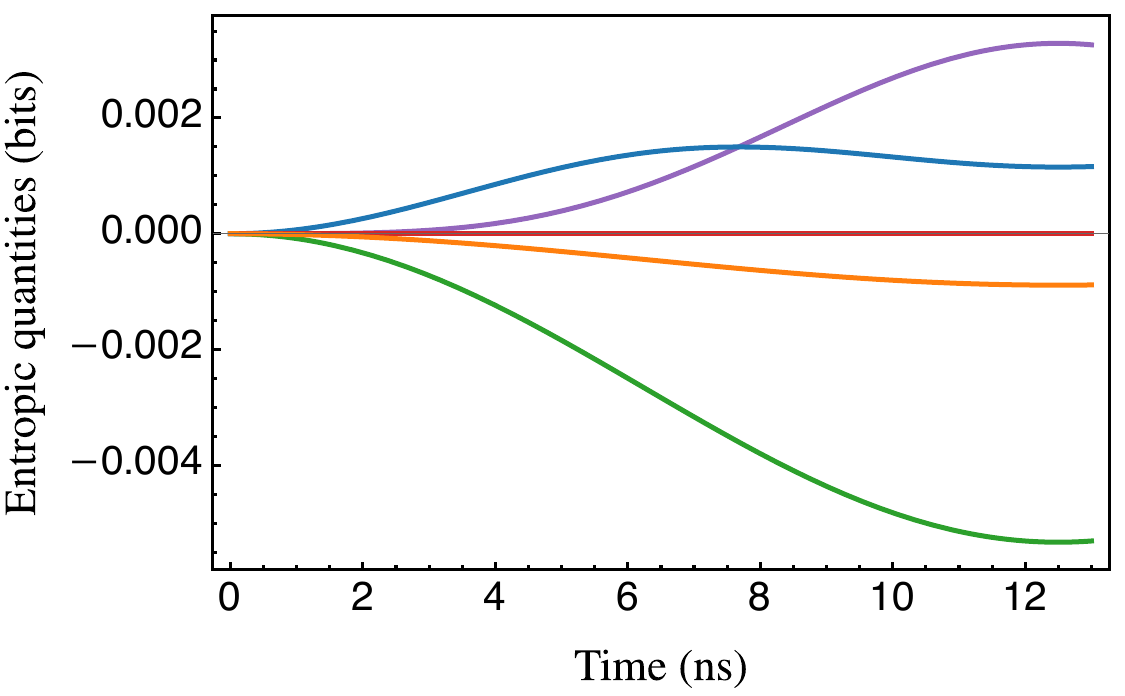}\label{triparitite_contour}
\put(22,55){\small{\textsf{(a)}}}
\end{overpic}
\end{minipage}
}
\hspace{-0.05\textwidth}
\subfigure{
\begin{minipage}[]{0.49\textwidth}	
\begin{overpic}[width=1\textwidth]{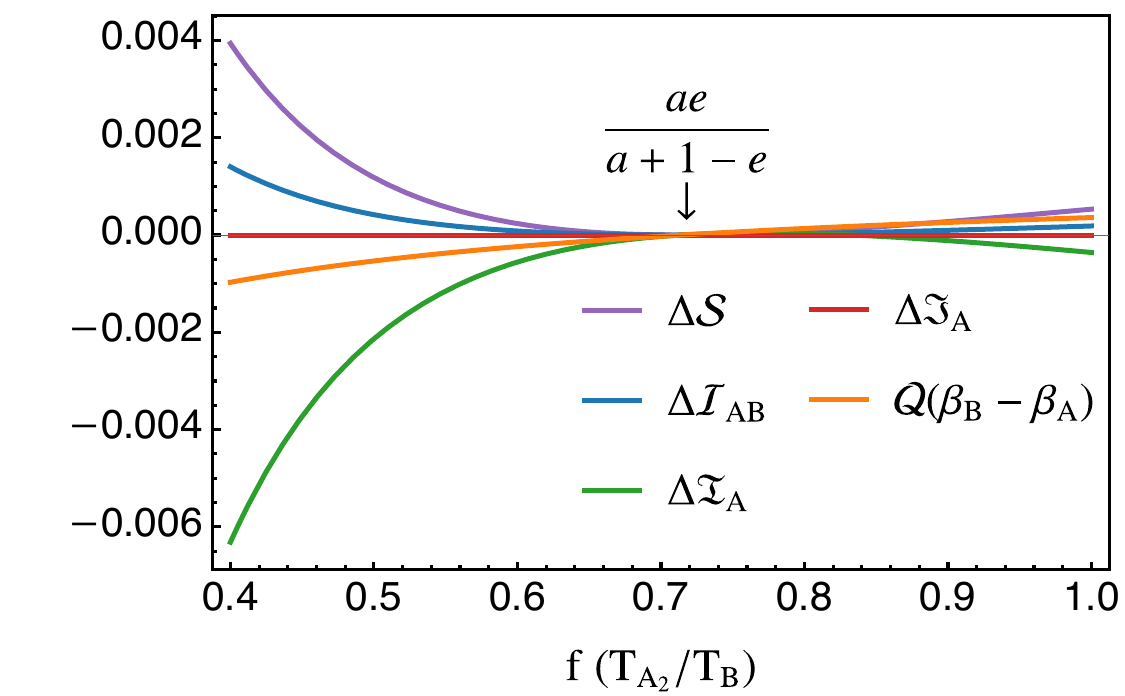}\label{triparitite_sustain}
\put(24,55){\small{\textsf{(b)}}}
\end{overpic}
\end{minipage}
}	
\caption{(a) Entropic quantities variation with interaction time for the AHT mechanism \eqref{AHT-mechanism-(I)}. (b) Entropic quantities vs. temperature of the qubit $A_2$ at $t=12\textrm{ns}$.
In both (a) and (b), we take $a=1$, $\omega_B=\omega_0$ with $\omega_0/ (2\pi) = 4 \textrm{GHz}$, $c=0.005 \omega_0$,
 $T_{A_1}=1.0 \textrm{K} \, (e=5/6)$, $T_B=1.2\textrm{K}$, and $ae/(a+1-e)=5/7 \doteq 0.71$. In (a) we take $T_{A_2}= 0.5 \textrm{K} \, (f=5/12 \doteq 0.42)$.
}\label{Example1}
\end{figure} 

With the above parameterization for initial state and Hamiltonian, it's straightforward to get each terms in Eq. (5), and  AHT region can be obtained by properly tune the parameters $a,e$ and $f$. 
After straightforward calculations, we find the variation of the states,
\begin{equation}
	\rho_{X}'(t)=\rho_X+\Delta\rho_X(t) ~   \mathrm{for} ~ X\in\{A_1,A_2,B\},  
\end{equation}
where 
\begin{equation}
\begin{aligned}
		&\Delta\rho_{A_1}(t)=
		\mathrm{diag}[-\Delta(t) , \Delta(t)],~\Delta\rho_{A_2}(t)=-\Delta\rho_{A_1}(t),
		~\Delta\rho_B(t)=-\Delta\rho_{A_1}(t) ,
\end{aligned}
\end{equation}
with 
\begin{equation}\label{DeltaTAndX}
	\Delta(t)\equiv\frac{ \left(e^{\frac{1+a}{e}\mathcal{X}}-e^{\frac{a}{f}\mathcal{X}} e^{\mathcal{X}}\right) \sin^2(ct) }{(1+e^{\frac{1+a}{e}\mathcal{X}})(1+e^{\frac{a}{f}\mathcal{X}})(1+e^{\mathcal{X}})},  ~ ~\mathcal{X}\equiv \frac{\hbar\omega_0}{k_\mathrm{B}T_B} .
\end{equation}
Whereupon for the terms in Eq. (5) in the main text, we get the variation of mutual information, 
\begin{equation}\label{(I)DeltaIAB}
\begin{aligned}
\Delta \mathcal{I}_{AB} = & -[p^{A_1}_0-\Delta(t)] \log [p^{A_1}_0-\Delta(t)]-[p^{A_1}_1 +\Delta(t)] \log [p^{A_1}_1+\Delta(t)] \\
		&  -[p^{A_2}_0 + \Delta(t)] \log [p^{A_2}_0 + \Delta(t)]-[p^{A_2}_1 - \Delta(t)] \log [p^{A_2}_1 - \Delta(t)] \\
		& -[p^B_0 + \Delta(t)] \log [p^B_0 + \Delta(t)]-[p^B_1 - \Delta(t)] \log [p^B_1 - \Delta(t)] \\
		& +  \sum_{i=1}^2 p^{A_1}_i \log p^{A_1}_i + p^{A_2}_i \log p^{A_2}_i + p^B_i \log p^B_i ,
\end{aligned}
\end{equation}
where the initial populations $p^{A_1}_0=\frac{e^{\frac{1+a}{e}\mathcal{X}}}{1-e^{\frac{1+a}{e}\mathcal{X}}}$, $p^{A_1}_1=\frac{1}{1-e^{\frac{1+a}{e}\mathcal{X}}}$, $p^{A_2}_0=\frac{e^{\frac{a}{f}\mathcal{X}}}{1-e^{\frac{a}{f}\mathcal{X}}}$, $p^{A_2}_1=\frac{1}{1-e^{\frac{a}{f}\mathcal{X}}}$, $p^{B}_0=\frac{e^{\mathcal{X}}}{1-e^{\mathcal{X}}}$, and $p^{B}_1=\frac{1}{1-e^{\mathcal{X}}}$, the  relative entropy,   
\begin{equation}\label{(I)DeltaREntropy}
\begin{aligned}
\Delta\mathcal{S} 
=& \left( \frac{1+a-e}{e}-\frac{a}{f} \right)\mathcal{X} \Delta(t) -  \Delta \mathcal{I}_{AB}  ,
\end{aligned}
\end{equation}
and the entropy induced by the temperature inhomogeneity, 
\begin{equation}\label{(I)DeltaTI}
	\Delta \mathfrak{T}_A = \left(\frac{1}{f}-\frac{1}{e}\right) a \mathcal{X}\Delta(t) \ . 
\end{equation}
Hence from Eqs. \eqref{(I)DeltaIAB}-\eqref{(I)DeltaTI},  we find the solution for the AHT mechanism   
\begin{equation}\label{AHT-mechanism-(I)}
	\Delta \mathcal{S}+\Delta \mathcal{I}_{A B} < - \Delta \mathfrak{T}_A  
\end{equation}  
is
\begin{equation}\label{ef-relation}
  0 < f	< \frac{a e}{1+a-e},  
\end{equation} 
which is independent of $c$, $T_B$, and $\omega_0$. 
From Eq. \eqref{ef-relation}, we see that only if the temperature of the qubit $A_2$ is low enough, that is,  $0< T_{A_2}<  [ ae/(1+a-e) ] T_B   < T_{A_1} < T_B$, the AHT mechanism  can be realized.  
Equivalently, we also have $1/\beta_B < \beta_{A_1}/\beta_B < [a/(1+a)] \beta_{A_2}/\beta_B + [1/(1+a) ] 1/\beta_B $ from Eqs. \eqref{parameterize-TA1TA2} and \eqref{ef-relation}, as appeared in the main text.  Moreover, from Eq. \eqref{DeltaTAndX}, the amount of  AHT gets maximum at the time points 
\begin{equation}
	T^{\max}_n=\frac{1}{c}\pi (n+\frac{1}{2}) ~  \textrm{with} ~ n\in \mathbb{Z}^+ .
\end{equation}

Fig. \ref{Example1} shows how entropic quantities behave with time and initial temperature inhomogeneity. In Fig. \ref{Example1} (b), the region for AHT mechanism is $0<T_{A_2}< [ae/(1+a-e)] T_B \doteq 0.86 \textrm{K} $, where $T_{A_2}=0.5 \textrm{K}$ gives the example of Fig. 1 (b) in the main text.

\vspace{5pt}

(II) \textbf{Intrasystem interaction.} To design an intrasystem interaction induced AHT  without any initial correlations and temperature inhomogeneity  ($\Delta \mathcal{I}_{AB} \geqslant 0$ and $\Delta \mathfrak{T}_A =0$), we also choose the product state \eqref{productstate} as the initial state. Here there is no intrasystem temperature inhomogeneity, that is,  $e=f$ in Eq. \eqref{parameterize-TA1TA2}. 
According to the Theorem 1, 3B interaction is necessary to realize the AHT since there is no initial quantum correlations among the qubits. Hence we parameterize a combination of 2B and 3B interaction from Eqs. \eqref{2B3BHAI} and \eqref{HI_23mixed}, which can written as 
\begin{equation}
\begin{aligned}
& \begin{aligned} H_I = & 2 \left\{ \mathcal{C} \left[ (2-r_1)\sigma^-_{A_1} \sigma^+_{A_2} \sigma^-_B - r_1 \sigma^+_{A_1} \sigma^-_{A_2} \sigma^-_B \right] + \textrm{h.c.} \right\}  + 2 \left[\mathcal{C} \, r_2 \, (\sigma^z_{A_2} \sigma^-_B - \sigma^z_{A_1} \sigma^-_B) + \textrm{h.c.} \right] , \end{aligned}  \\
& H_{A I}=\omega_0 \left(r_2 \sigma_{A_1}^{+} \sigma_{A_2}^{-}+\text {h.c. }\right)  ,
\end{aligned}
\end{equation}
where $\omega_{A_1} / r_1=\omega_{A_2}=\omega_B \equiv \omega_0$, $r_1 \in (0, 2]$, $r_2 = \pm \frac{1}{2} \sqrt{r_1 (2 - r_1) } \in [-1/2, 1/2] $, and  $\mathcal{C}=(c^{xyy}+c^{xyx} i)$ with $c^{xyy,xyx}$ are real. 

 With the above parameterization, after some straightforward calculations,  one will find the solution  for the AHT mechanism 
 \begin{equation}\label{AHT-mechanism-(II)}
 	\Delta \mathcal{S}+\Delta \mathcal{I}_{A B} < - \Delta \mathfrak{I}_A 
 \end{equation}
is $\Theta >0 $, where 
 \begin{equation}
 	\Theta \equiv (r_1-2) \left[ \exp{ \left(\frac{1}{e} \mathcal{X} \right) } - \exp \left( \frac{ e + r_1 }{e}  \mathcal{X} \right) \right]+  r_1 \left[ \exp \left( \frac{ e+1  }{e}  \mathcal{X} \right) -  \exp \left( \frac{ r_1 }{e} \mathcal{X} \right) \right] .
 \end{equation}
 
\begin{figure}[htb]
\centering
\subfigure{
\begin{minipage}[]{0.49\textwidth}
\begin{overpic}[width=1\textwidth]{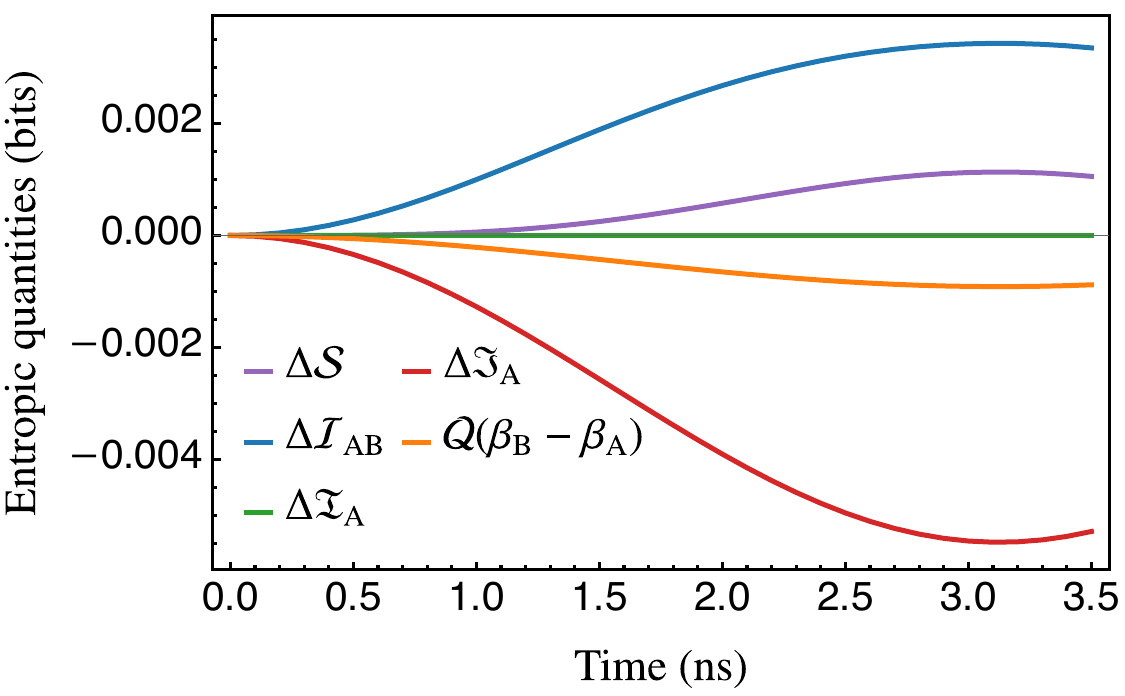}\label{triparitite_contour}
 \put(22,55){\small{\textsf{(a)}}}
\end{overpic}
\end{minipage}
}
\hspace{-0.05\textwidth}
\subfigure{
\begin{minipage}[]{0.49\textwidth}	
\begin{overpic}[width=1\textwidth]{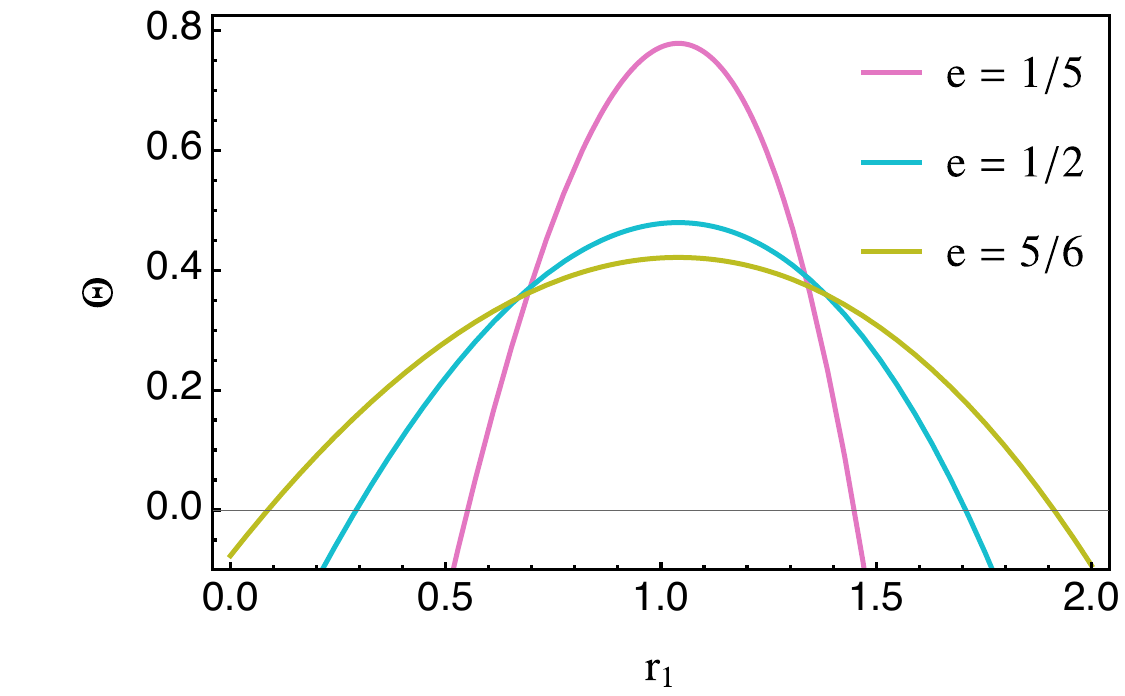}\label{triparitite_sustain}
 \put(22,55){\small{\textsf{(b)}}}
\end{overpic}
\end{minipage}
}	
\caption{(a) Entropic quantities variation with interaction time for the AHT mechanism  \eqref{AHT-mechanism-(II)}.  Here we take $T_B=1.2 \textrm{K}$, $T_{A_{1,2}}= 1.0 \textrm{K} \, (e=f=5/6)$, $\omega_B=\omega_0$ with $\omega_0/ (2\pi) = 4 \textrm{GHz}$, $r_1=1$, and $\mathcal{C}= 0.005 \omega_0$.
(b)  $\Theta$ as a function of $r_1$  ($\mathcal{X} \doteq 0.16$).}
\label{Example2}
\end{figure} 

Fig. \ref{Example2} (a) shows how the entropic quantities  behave with interaction time for this AHT mechanism. Fig. \ref{Example2} (b) shows $\Theta$ as a function of $r_1$, where the region for the AHT mechanism decreases as the temperature difference between the systems A and B increases ($e$ decreases). Note when $r_1=1$ ($r_2=1/2$ accordingly), we aways have  $\Theta>0$ for $e\in (0,1)$, which is the case of the example (II) and Fig. 1 (c) in the main text.

\vspace{5pt}

(III) \textbf{Initial state correlation.} Here to design an initial intrasystem correlation induced AHT without temperature inhomogeneity and intrasystem interaction ($\Delta \mathcal{I}_{A:B} \geqslant 0$, $\Delta \mathfrak{T}_A =0 $,  and $\Delta \mathfrak{I}_A =0 $), we whereupon parameterize the initial state as
\begin{equation}
	\left[ \rho_{A_1} \otimes \rho_{A_2}+ \Lambda  \left(\sigma_{A_1}^{+} \otimes  \sigma_{A_2}^{-} +  \textrm{h.c.} \right) \right] \otimes \rho_B , 
\end{equation} in which  the intrasystem  correlation parameter 
 \begin{equation}
 	| \Lambda |  \leqslant \Lambda_{\max} \equiv 1/4  \textrm{sech}^2 [\mathcal{X}/(2e)] 
 \end{equation} 
 to make the state to be semi-positive.  The vanishing  intrasystem temperature inhomogeneity requires  $e=f$ in Eq. \eqref{parameterize-TA1TA2}. We also set $H_{AI}=0$ to make sure $\Delta \mathfrak{I}_A =0 $.
From the derivation  of  Theorem 1 in section IV, we observe  2B interaction is sufficient to realize AHT for an initially  correlated state. Hence we  parameterize the interaction to be 
\begin{equation}
	H_I= \eta  \omega_0\left(\sigma_{A_1}^{-} \sigma_B^{+}+\sigma_{A_2}^{-} \sigma_B^{+}+\text {h.c. }\right)  
\end{equation}
according to Eq. (8). Accordingly, the qubit frequencies satisfy $\omega_{A_{1,2}}=\omega_B\equiv \omega_0$.

\begin{figure}[t]
\centering
\subfigure{
\begin{minipage}[]{0.49\textwidth}
\begin{overpic}[width=1\textwidth]{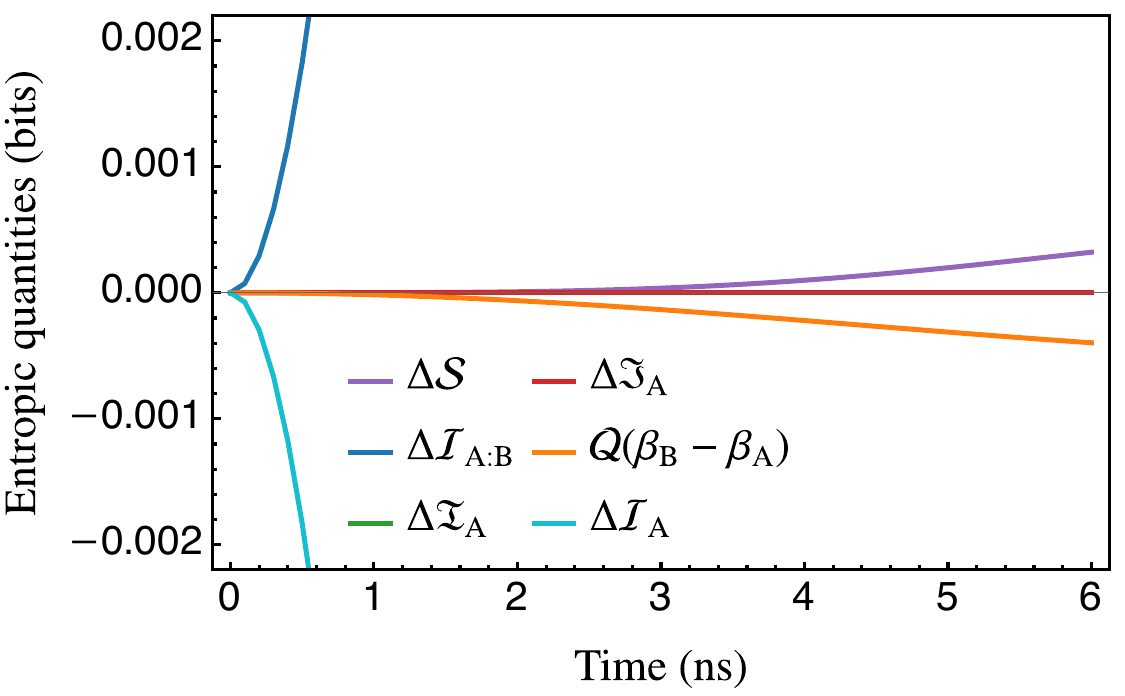}\label{triparitite_contour}
 \put(21,55){\small{\textsf{(a)}}}
\end{overpic}
\end{minipage}
}
\hspace{-0.05\textwidth}
\subfigure{
\begin{minipage}[]{0.49\textwidth}	
\begin{overpic}[width=1\textwidth]{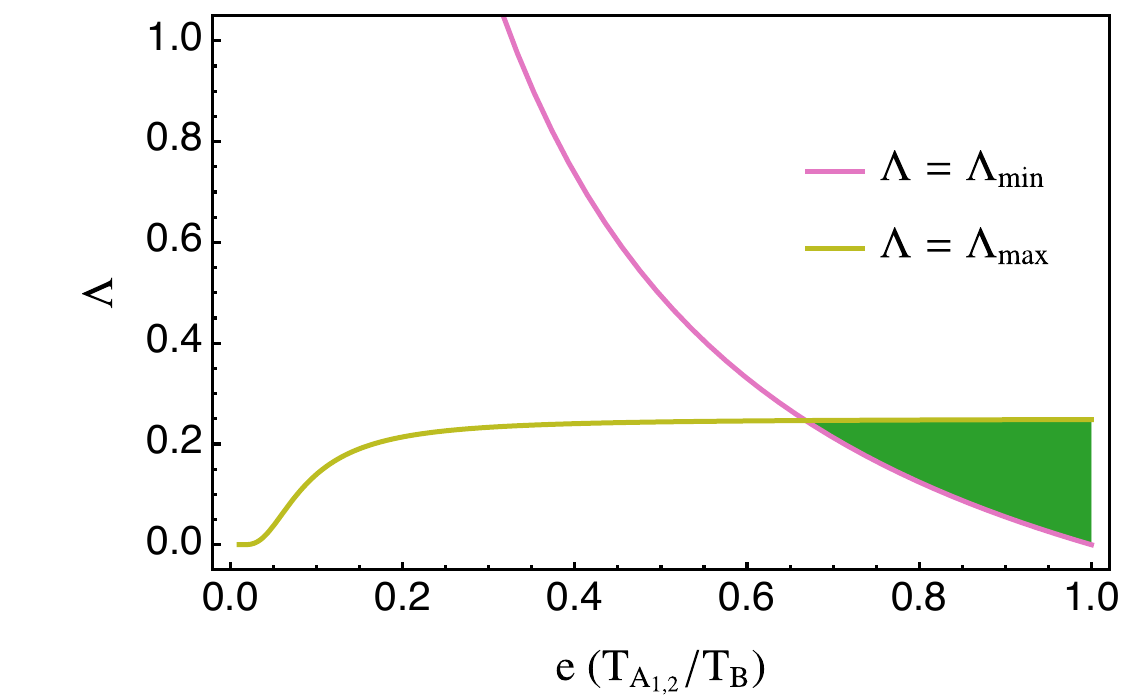}\label{triparitite_sustain}
 \put(22,55){\small{\textsf{(b)}}}
\end{overpic}
\end{minipage}
}	
\caption{ (a) Entropic quantities variation with interaction time for the AHT mechanism  \eqref{AHT-mechanism-(III)}. Here we take $T_B=1.2 \textrm{K}$, $T_{A_{1,2}}= 1.0 \textrm{K} \, (e=f=5/6)$, $\omega_{A_{1,2}}=\omega_B=\omega_0$ with $\omega_0/ (2\pi)  = 4 \textrm{GHz}$, $\Lambda=6/25=0.24$ ($\Lambda_{\min} \doteq 0.10, \Lambda_{\max} \doteq 0.25$), and $\eta=1/200$.
(b) Diagram for the region of AHT mechanism (green):   $\Lambda_{\min} < \Lambda < \Lambda_{\max} $. Here $\mathcal{X} \doteq 0.16$ ($T_B=1.2\textrm{K}$).
}
\label{Example3}
\end{figure}
 
With the above initial state and Hamiltonian, it is easy to get each terms in Eq. (5), and 
here we find the solution  for the AHT mechanism 
\begin{equation}\label{AHT-mechanism-(III)}
		\Delta \mathcal{S} + \Delta \mathcal{I}_{A:B}  <  - \Delta \mathcal{I}_{A} 
\end{equation} 
is 
\begin{equation}
	\Lambda > \Lambda_{\min} \equiv \frac{\exp(\mathcal{X}/e)-\exp(\mathcal{X})}{\left[ \exp(\mathcal{X}) - 1 \right] \left[ \exp(\mathcal{X}/e) + 1 \right]}  . 
\end{equation}
Fig. \ref{Example3} (b) shows the region for this AHT mechanism. One should note  only if  $\Lambda_{\min} <  \Lambda_{\max}$ it is possible to realize the AHT mechanism.
Fig. \ref{Example3} (a) shows how entropic quantities variate with interaction time, which is of the same parameter setting of the example (III) and Fig. 1 (d) in the main text.

\section{Energy balance equation for broken HTC}

If the variation of intersystem interaction energy $\Delta \langle H^I_{AB} \rangle $ is not zero, from energy conservation for the whole system,  we have $\Delta \langle  H_A \rangle  = - \Delta \langle H_B \rangle  - \Delta \langle  H^I_{AB} \rangle $ instead of $\Delta \langle H_A \rangle = - \Delta \langle  H_B \rangle $.  Referring  to the derivation in section II, one readily  find that the Eq. (5) in the main text becomes 
\begin{equation}\label{AHT-mechanism-BreakHTC}
	(\beta_B-\beta_A) \Delta \langle H_B \rangle = \Delta \mathcal{S}+\Delta \mathcal{I}_{A B}+\Delta \mathfrak{T}_A + \Delta \mathfrak{I}_{AB},  \ \ \ \Delta  \mathfrak{I}_{AB} = \Delta \mathfrak{I}_A +  \Delta \mathfrak{I}_{A:B}, 
\end{equation}  
where   $\Delta \mathfrak{I}_{A:B}\equiv \beta_{A} \Delta  \langle H^I_{AB} \rangle $ and $\Delta \mathfrak{I}_{AB}$  denote  the entropic quantities induced by variation of interaction energy  between the systems and of the whole system, respectively.  
Here Eq. \eqref{AHT-mechanism-BreakHTC} holds for an arbitrary interaction Hamiltonian, and HTC is broken if $\Delta \mathfrak{I}_{A:B} \neq 0$.


\begin{figure}[htb]
\centering
\subfigure{
\begin{minipage}[]{0.49\textwidth}
\begin{overpic}[width=1\textwidth]{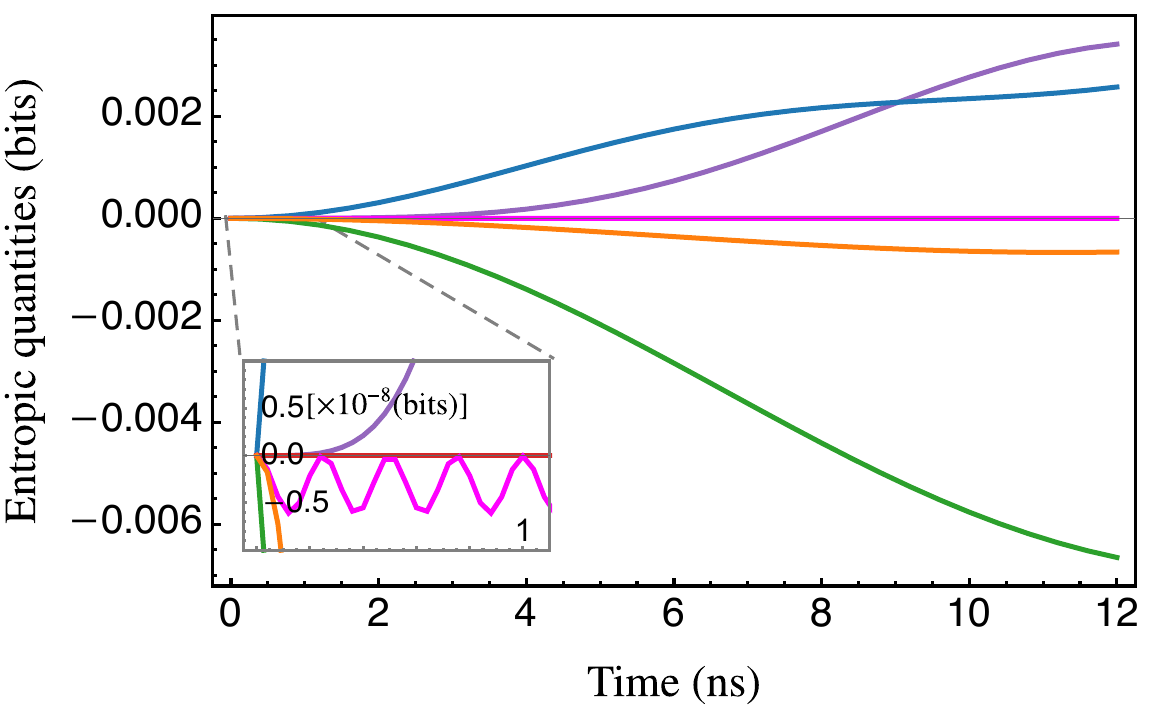}\label{triparitite_contour}
 \put(21,55){\small{\textsf{(a)}}}
\end{overpic}
\end{minipage}
}
\hspace{-0.05\textwidth}
\subfigure{
\begin{minipage}[]{0.49\textwidth}	
\begin{overpic}[width=1\textwidth]{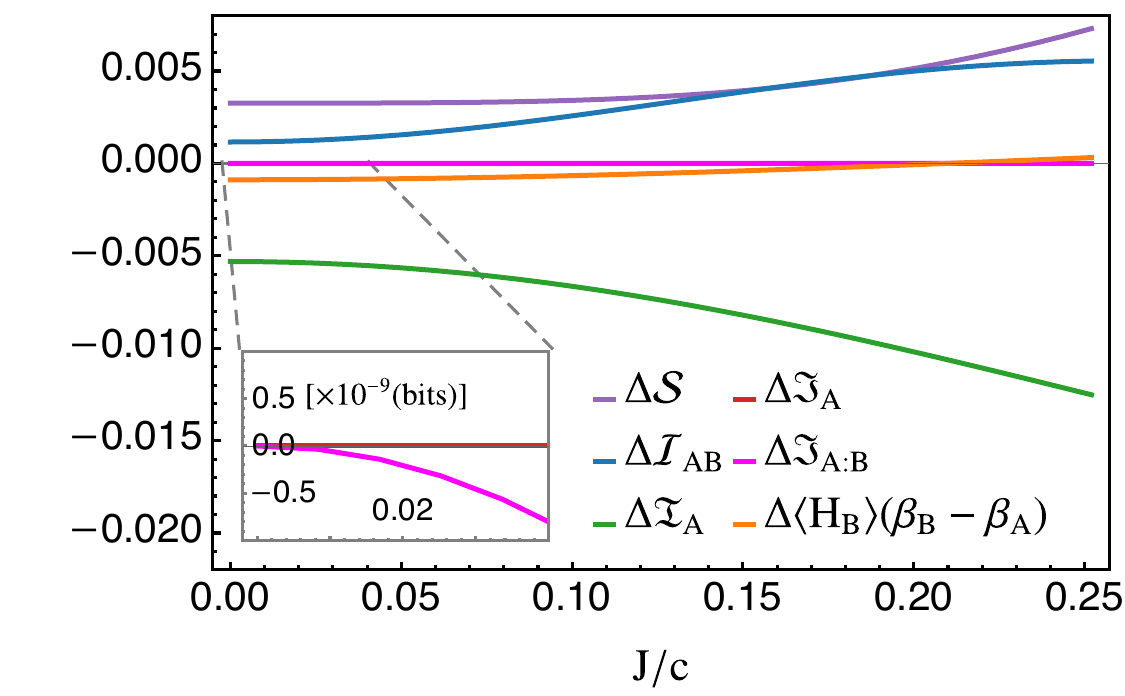}\label{triparitite_sustain}
 \put(22,55){\small{\textsf{(b)}}}
\end{overpic}
\end{minipage}
}	
\caption{(a) Entropic quantities variation with time under the combination of interactions \eqref{HI-(I)} and  \eqref{HI-(1)-MixedBreakHTC} at $J/c=0.1$.
 Other parameters setting are the same as those in Fig. \ref{Example1} (a).
(b) Entropic quantities variation vs. $J/c$ at $t=12 \textrm{ns}$. }
\label{MixedBreakHTC}
\end{figure} 


\begin{figure}[htb]
\centering
\subfigure{
\begin{minipage}[]{0.49\textwidth}
\begin{overpic}[width=1\textwidth]{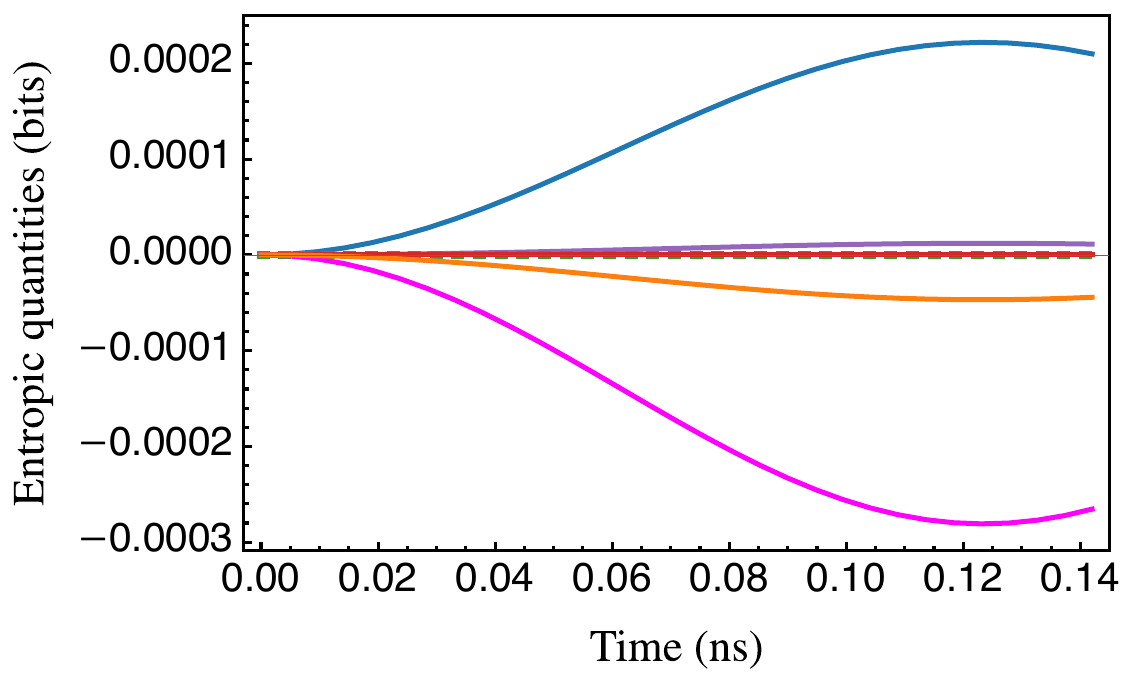}\label{triparitite_contour}
 \put(24,53){\small{\textsf{(a)}}}
\end{overpic}
\end{minipage}
}
\hspace{-0.03\textwidth}
\subfigure{
\begin{minipage}[]{0.49\textwidth}	
\begin{overpic}[width=1\textwidth]{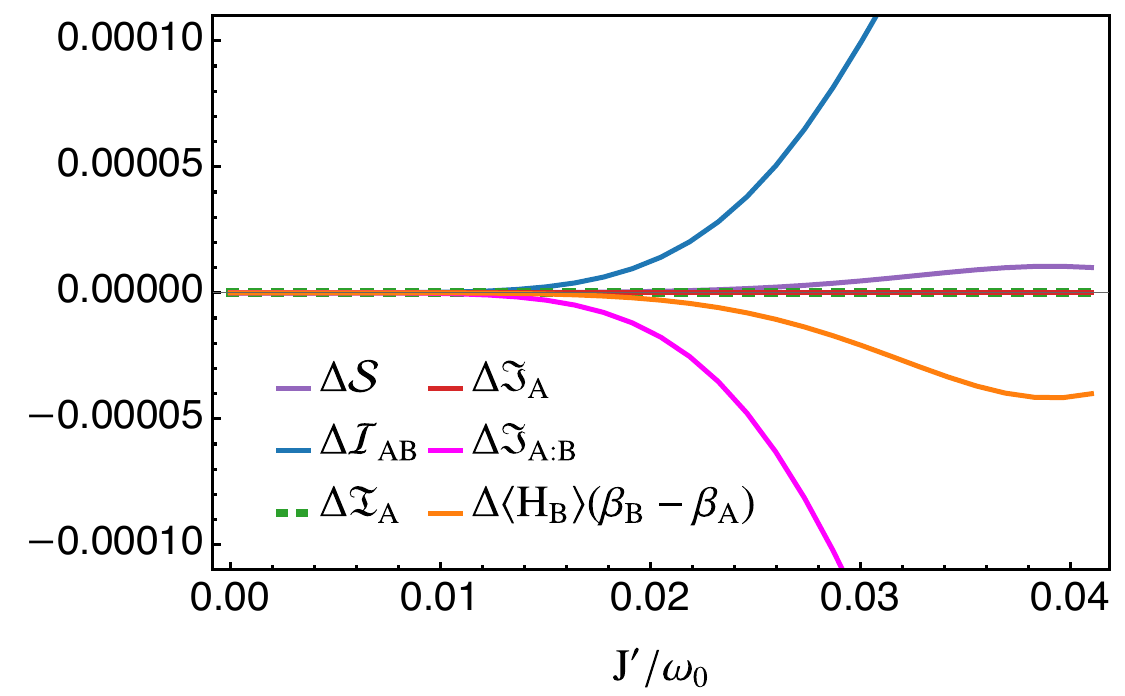}\label{triparitite_sustain}
 \put(22,55){\small{\textsf{(b)}}}
\end{overpic}
\end{minipage}
}	
\caption{(a) Entropic quantities variation with interaction time under the interaction Eq. \eqref{HI-(1)-BreakHTC}  that breaks HTC at  $J'/\omega_0=0.04$. Here we take $T_B=1.2 \textrm{K}$, $T_{A_{1,2}}= 1.0 \textrm{K}$, and other settings are the same as those in Fig. \ref{Example1} (a).   (b) Entropic quantities variation vs. coupling strength  $J'/\omega_0$ at $t=12 \textrm{ns}$.
}
\label{BreakHTC}
\end{figure} 

Fig. \ref{MixedBreakHTC} shows the entropic quantities variation when interaction in the example (I) is mixed with 
 the interaction
\begin{equation}\label{HI-(1)-MixedBreakHTC}
H^I_b =	J \sum_{ <u,v>} \sum_{a=x}^z  \sigma_u^a  \sigma_v^a, 
\end{equation}
which violates HTC. We can see that when $J/c$ is small [$\lesssim 0.21$ in Fig. \ref{MixedBreakHTC} (b)], 
energy still  flows  out from  the system B although its temperature is higher, which shows the robustness of AHT as mentioned on Page 4 of the main text.  
In addition, the variation of intersystem coupling energy does can draw out  energy from the higher temperature system.  Fig. \ref{BreakHTC} shows the entropic quantities variations under the HTC breaking interaction     
\begin{equation}\label{HI-(1)-BreakHTC}
	H^I_{b'}= J' \left( \sigma_{A_1}^z \sigma_B^x + \sigma_{A_2}^z \sigma_B^x \right),
\end{equation}
where $J'$ is coupling strength.  In Fig. \ref{BreakHTC},  there is no  intarsystem interaction and temperature inhomogeneity, hence  $\Delta \mathfrak{T}_A=\Delta \mathfrak{I}_A = 0$ over the evolution. 
Energy is  drawn out from the system B since $-\Delta \mathfrak{I}_{A:B} >  \Delta \mathcal{S} + \Delta \mathcal{I}_{AB} >0$   from Eq. \eqref{AHT-mechanism-BreakHTC}.

\end{document}